\documentclass[12pt, draftclsnofoot, onecolumn]{IEEEtran}
\usepackage{graphicx}
\usepackage{subfigure}
\usepackage{booktabs}
\usepackage{multirow}
\usepackage{verbatim}
\usepackage{subeqnarray}
\usepackage{bm}
\usepackage[noend]{algpseudocode}
\usepackage{algorithmicx,algorithm}
\usepackage{cite}%
\usepackage[
            linkcolor=red,
            anchorcolor=blue,
            citecolor=green]{hyperref}
\usepackage{amsmath}
\usepackage{amssymb}
\usepackage{color}
\usepackage{amsmath,bm}
\usepackage[amsmath,thmmarks]{ntheorem}
\theoremseparator{:}

\ifCLASSINFOpdf

\else

\fi

\begin{document}
\title{LEOS-assisted Inter-GEOS Communication via Distributed-storage Coding}

\author{Haisheng Xu,~\IEEEmembership{Member,~IEEE},
~Chen Gong,~\IEEEmembership{Member,~IEEE},
~Haipeng Yao,~\IEEEmembership{Member,~IEEE},
~and Xiaodong Wang,~\IEEEmembership{Fellow,~IEEE}

\thanks{H. Xu and X. Wang are with the Department of Electrical Engineering, Columbia University, New York, NY 10027, USA. (e-mail: hx2219@columbia.edu, wangx@ee.columbia.edu). 
\par C. Gong is with the Key Laboratory of Wireless Optical
Communications, Chinese Academy of Sciences, School of Information
Science and Technology, University of Science and Technology
of China, Hefei 230027, China (e-mail: cgong821@ustc.edu.cn). \par H. Yao is with the State Key Laboratory of Networking and Switching Technology, Beijing University of Posts and Telecommunications, Beijing 100876, China (e-mail: yaohaipeng@bupt.edu.cn).}
}

\markboth{ }%
{Shell \MakeLowercase{\textit{et al.}}: Bare Demo of IEEEtran.cls for Journals}

\IEEEtitleabstractindextext{
\begin{abstract}
We consider a space communication network consisting of Geosynchronous Earth Orbit satellites (GEOSs) and Low Earth Orbit satellites (LEOSs). In case of no direct communication link between two GEOSs, the data exchange between them is through relay by the LEOSs. In particular, the source GEOS sends coded data to multiple LEOSs based on the distributed storage framework. The destination GEOS then retrieves certain amount of data from each LEOS for data reconstruction. For the GEOS-LEOS downlink, a regenerating-code-based transmission scheme is optimized to guarantee data reconstructability, where the transmission power allocation to the LEOSs is proposed to minimize the total transmission energy. We also consider the power allocation to minimize the total transmission time given the total transmission energy. For the LEOS-GEOS uplink,
a flexible partial-downloading coding transmission scheme is proposed to guarantee data reconstructability, where the joint uploaded-data size and power allocations are proposed to minimize the total transmission energy or the total transmission time. Extensive simulation results are presented to evaluate the proposed algorithms and show that regenerating code can achieve lower transmission energy and shorter transmission time for data regeneration than conventional maximum-distance separable (MDS) code.
\end{abstract}

\begin{IEEEkeywords}
 Inter-GEOS communication, relay, distributed storage, regenerating codes, resource allocation, outer approximation (OA) method.
\end{IEEEkeywords}}

\maketitle

\IEEEdisplaynontitleabstractindextext
\IEEEpeerreviewmaketitle

\section{Introduction}
\par Since the first artificial satellite was launched in the 1950s, satellite communications have experienced rapid development with the explosive growth in the number of Earth-orbiting satellites. Satellites have played a significant and important role in both civil and military applications such as seamless communications, remote sensing and global reconnaissances\cite{Chitre1999,Evans2005,Kandeepan2010,Xu2015c,Du2016a}. As modern satellites become more and more powerful in acquiring, processing and storing large amount of data, and in the meantime, they become more and more interconnected, efficient data exchange among these satellites becomes a pressing challenge.
\par Satellites can be classified according to their orbital altitudes as Low Earth Orbit (LEO), Medium Earth Orbit (MEO) and Geostationary Earth Orbit (GEO) systems\cite{Sharma2016}. For data transmission to the ground stations (GSs), MEO/LEO satellites (MEOSs/LEOSs) have lower propagation attenuation and power consumption compared with the GEO satellites (GEOSs) \cite{Yang2012}. However, MEOSs/LEOSs cannot provide consistent coverage due to the fast movements of footprints and limited illuminating region on the Earth; while GEOSs on the geo-synchronous orbit can provide consistent and reliable coverages, and thus can serve as data centers capable of storing, processing and distributing data \cite{yao2018,Evans2005,Du2016a}. Therefore, it is easier to establish a reliable communication link between the GEOSs and GSs or between the GEOSs and MEOSs/LEOSs than do it between the MEOSs/LEOSs and GSs. As far as we know, resource allocations over the GEOS-GS link for data transmissions have been extensively studied \cite{Choi2005,Zheng2012,Vassaki2013,Lagunas2015,Aravanis2015,Du2017} and the GEOS-GS link using two GEOSs has also been considered \cite{Sharma2015,Jin2016}. However, to the best of our knowledge, little literature refers to data transmission issues over space links since big space data can be generated from Earth observations, deep space explorations and human information \& network businesses and have become a pressing challenge and attracted great attentions by NASA and other organizations all over the world \cite{Earley2016,Chen2018}. As the GEOSs can collect space data from near Earth spacecrafts, deep space spacecrafts, and even terrestrial publishers and these data finally need to be sent to Earth for
human service, we consider the following scenario in this paper for space-link-based data transmissions: in a space network, there are only two high-cost GEOSs serving as data centers due to limited launching budget; one GEOS transmits some important data but does not cover the target GSs, which are instead covered by the other GEOS without the data. Due to geopolitical limitations, the GEOS with the data cannot communicate to the target GSs via the GSs in its coverage area, or cannot directly communicate to the other GEOS due to the Earth blockage. In such a scenario, we consider the data transmission between the two GEOSs assisted by relaying through low-cost MEOSs/LEOSs since such a way does not have any geopolitical issues. The data transmissions under consideration can be viewed as a relay for data transmissions between the source GEOS and target GSs.
\par On the other hand, distributed storage techniques have received significant recent interest as they can provide high reliabilities for stored data over long periods of time. These reliabilities are mainly attributed to the redundancy in storage and the basic idea is to distribute data to different nodes through maximum-distance separable (MDS) or regeneration coding such that the original data can be recovered and any failed node can be repaired by downloading data from a subset of these nodes\cite{Lacan2004,Cadambe2013,Suh2011,Dimakis2010,Calis2016,Shah2010,Gong2012}. Although both MDS and regenerating codes can achieve optimal reconstruction, MDS codes usually treat the content stored in each node as a single data/symbol over a finite filed $\mathbb{F}$, resulting in that the data amount of the whole original data should be downloaded for failed node repair \cite{Lacan2004,Suh2011,Cadambe2013}. However, the regenerating codes treat the content as being comprised of several subdata/symbols over $\mathbb{F}$ and thus can realize efficient failed node repair with lower data amount\cite{Dimakis2010,Rashmi2011}. With the developments of low-cost high-volume storage devices and miniaturized payloads, installing distributed space storage and transponder capability on low-orbiting satellites becomes feasible. These satellites can then form a distributed storage system in space to assist the communications between GEOSs. And the redundancy of the distributed stored data can increase the robustness of the storage system, which raises the questions of resource allocation and transmission protocol design to effectively save transmission energy or time.
\par In this paper, we propose efficient transmission schemes for such inter-GEOSs communications assisted by LEOSs using distributed-storage coding, where the direct link between the two GEOSs does not exist. The source GEOS sends coded data to a group of LEOSs, which are then retrieved by the destination GEOS for data reconstruction. The transmission system consists of two types of data-links: GEOS-LEOS downlink and LEOS-GEOS uplink. For the GEOS-LEOS downlink, a regenerating-code-based \cite{Dimakis2010,Rashmi2011} transmission scheme is proposed to guarantee the data reconstructability, and transmission power allocation to different LEOSs is optimized to minimize the total transmission energy or the total transmission time. For the LEOS-GEOS uplink, a flexible partial-downloading transmission scheme is proposed to guarantee the data reconstructability, and a joint uploaded-data size and power allocation is proposed to minimize the total transmission energy or transmission time. Finally, extensive simulation results are presented to evaluate the proposed algorithms and show that regenerating code can achieve lower transmission energy and shorter transmission time for data regeneration than conventional maximum-distance separable (MDS) code.
\par The remainder of this paper is organized as follows. In Section \ref{sec:ProbForm}, system models are presented and problem formulations are given. In Sections \ref{sec:Solution_GEO-LEO} and \ref{sec:Solution_LEO-GEO}, algorithms are developed to solve the resource allocation problems for the GEOS-LEOS downlink and the LEOS-GEOS uplink, respectively. In Section \ref{sec:RRforFailedLEOS}, the issues of resource allocation for failed LEOS repair are preliminarily discussed. In Section \ref{sec:NumResult}, simulation results are presented. Finally, conclusions are drawn in Section \ref{sec:Conclusion}.
\begin{figure}
  \centering
  \includegraphics[scale=0.9]{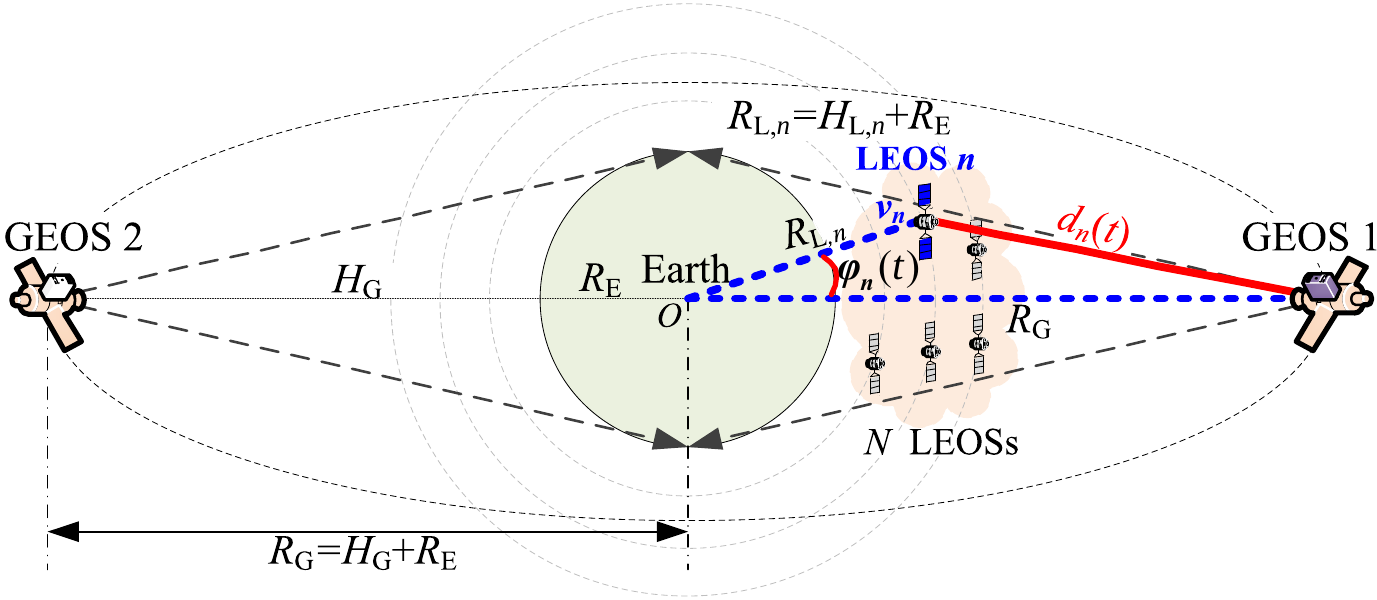}
 \caption{The schematic diagram of an inter-GEOS communication network with 2 GEOSs and $N$ distributed storage LEOSs.}\label{fig:Fig_SatNetwworkDemo}
  \end{figure}
\section{System Descriptions and Problem Formulation}\label{sec:ProbForm}
\subsection{System Model}
 We consider a system consisting of 2 GEOSs and $N$ circularly orbited LEOSs denoted as $\mathcal{N}=\{1,2,...,N\}$, where the $N$ LEOSs can operate in different orbital planes but, for simplicity, we draw them in a two-dimensional (2D) plane as shown in Fig. \ref{fig:Fig_SatNetwworkDemo}. The GEOSs serve as data centers capable of storing, processing and distributing data. However, there is no direct link between the two GEOSs.
Instead, the LEOSs serve as distributed space storage and transponder nodes, which can establish reliable communication links with the GEOSs when entering their coverage areas.
 \par As shown in Fig. \ref{fig:Fig_SatNetwworkDemo}, let $H_{\mathrm{G}}$ denote the altitude of the GEOSs, $H_{\mathrm{L},n}$ denote the altitude of LEOS $n$ with velocity $v_{n}$ for $n\in\mathcal{N}$ and $R_{\mathrm{E}}$ denote the radius of the Earth. Define $\varphi_{n}(t)$ as the instantaneous rotation angle of LEOS $n$ at time $t$, which is measured by the angle between LEOS $n$ and GEOS $1/2$ as seen from the center (point $O$) of the Earth. Since $\varphi_{n}(t)$ is unique for LEOS $n$ and can be easily calculated once the orbital parameters of LEOS $n$ are determined, we can use polar coordinates to represent all the satellite locations without loss of generality, i.e., $(R_{\mathrm{G}},0)$ for GEOS $1$, $(R_{\mathrm{G}},\pi)$ for GEOS $2$ and $(R_{\mathrm{L},n}, \varphi_{n}(t))$ for LEOS $n$ at time $t$. As point $O$, LEOS $n$ and GEOS $1/2$ define a 2D plane, the instantaneous distance between LEOS $n$ and GEOS $1/2$ can then be given by
\begin{equation}\label{equ:GEO-LEOlinkDistance}
\begin{split}
  d_{n}(t)&=\sqrt{R^{2}_{\mathrm{G}}+R_{\mathrm{L},n}^{2}-2R_{\mathrm{G}}R_{\mathrm{L},n}\cos(\varphi_{n}(t))},
 \end{split}
\end{equation}
which is obtained by the law of Cosines according to Fig. \ref{fig:Fig_SatNetwworkDemo}.
\begin{figure}[ht!]
  \centering
  \includegraphics[scale=1]{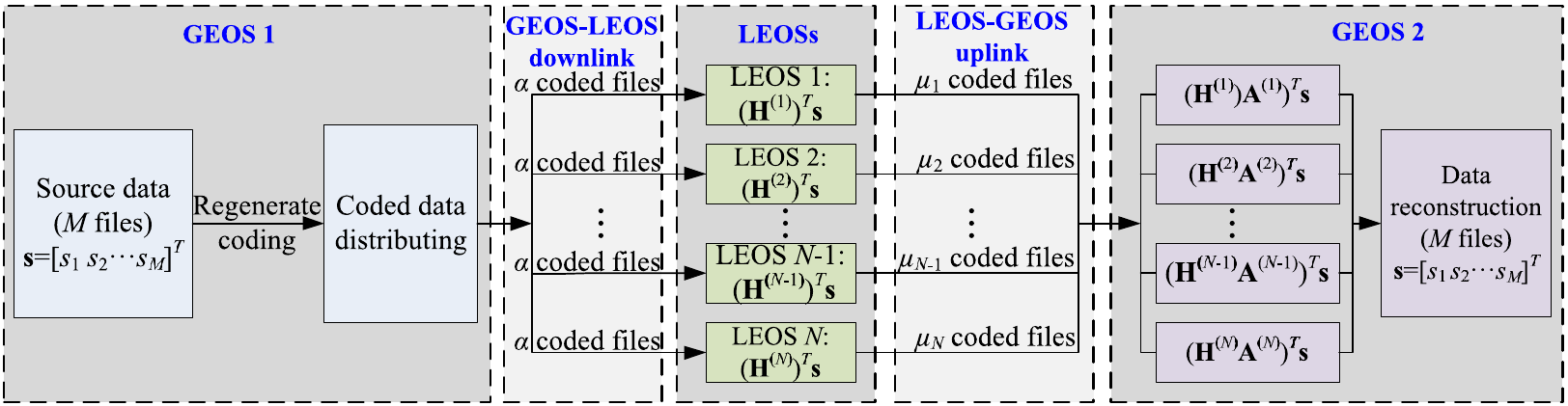}
  \caption{Diagram of data transmission for an inter-GEOS communication network.}\label{fig:Fig_DiagramDataTransmission}
  \end{figure}
\par We assume that GEOS 2 needs to obtain a source data of $M$ files $\mathbf{s}=[s_{1}~s_{2} \cdots s_{M}]^{T}$ stored on
GEOS 1 with the assistance of the $N$ LEOSs, where the superscript $^{T}$ denotes the transpose operator and each file is a symbol packet consisting of $u$ bits. Then the data transmission
is composed of two stages: the GEOS-LEOS downlink and the LEOS-GEOS uplink, as shown in Fig. \ref{fig:Fig_DiagramDataTransmission}. We employ an $(M,N,K,D,\alpha,\beta)$ regenerating
coding scheme \cite{Dimakis2010,Rashmi2011} for the first stage such that each of the $N$ LEOSs receives $\alpha$ coded files and the original $M$ files can be reconstructed by downloading $\alpha$ files each from any $K$ LEOSs (termed as $(\alpha,K)$-reconstructability); moreover, if the stored files on one LEOS get lost, the lost files can be regenerated by downloading $\beta$ files each from any other $D$ LEOSs. Under the regenerating-code-based distributed storage, we then employ a flexible downloading scheme \cite{Shah2010,Gong2012} for the second stage such that GEOS 2 can reconstruct the original $M$ files by downloading $\mu_{n}\leq\alpha$ coded files from LEOS $n$ for all $n\in\mathcal{N}$ (termed as $\bm\mu$-reconstructability). The detailed transmission schemes for the two stages are described as follows.
\begin{itemize}
  \item GEOS-LEOS downlink: When the $N$ LEOSs enter the coverage area of GEOS 1, each LEOS receives $\alpha$ linearly coded files of the data from GEOS 1. In particular, the received files of LEOS $n$ is given by
\begin{equation}\label{equ:Storagedata}
\begin{split}
 \mathbf{m}^{(n)}&=[m_{1}^{(n)}~m_{2}^{(n)}\cdots m_{\alpha}^{(n)}]^{T}\\
&=[\mathbf{h}_{1}^{(n)}~\mathbf{h}_{2}^{(n)}~\cdots~\mathbf{h}_{\alpha}^{(n)}]^{T}\mathbf{s}\\
&\triangleq \left(\mathbf{H}^{(n)}\right)^{T}\mathbf{s},
\end{split}
\end{equation}
where $\mathbf{h}_{i}^{(n)}=[h_{i1}^{(n)}~h_{i2}^{(n)}\cdots h_{iM}^{(n)}]^{T}$, $1\leq i\leq \alpha$, and $\mathbf{H}^{(n)}$,
$n\in\mathcal{N}$, are the encoding matrices of sizes $M\times \alpha$ known to both GEOSs. Note that for the $(M,N,K,D,\alpha,\beta)$ regenerating code, parameters $\alpha$ and $\gamma\triangleq D\beta$ are called the storage capacity and the repair bandwidth, respectively, and the code parameters should satisfy
      \begin{equation}\label{equ:AlphaReconstruction}
        K\leq D\leq N-1,~M\leq\sum_{i=0}^{K-1}\min\left\{\alpha,(D-i)\beta\right\}.
      \end{equation}
Two types of optimal operating conditions are usually of interest. The condition for the minimum storage regeneration (MSR) point is given by
\begin{equation}\label{equ:MSR_GEO-LEO}
 \alpha=\frac{M}{K},~\gamma=D\beta=\frac{MD}{(D-K+1)K};
\end{equation}
and the condition for the minimum bandwidth regeneration (MBR) point is given by
\begin{equation}\label{equ:MBR_GEO-LEO}
 \alpha=\frac{2MD}{2KD-K^{2}+K},~\gamma=D\beta=\frac{2MD}{2KD-K^2+K}.
\end{equation}
 \item LEOS-GEOS uplink: When the
LEOSs enter the coverage area of GEOS 2, LEOS $n$ transmits $\mu_{n}\leq\alpha$ coded files
to GEOS 2 in the form of $\left(\mathbf{A}^{(n)}\right)^{T}\mathbf{m}^{(n)}$,
where $\mathbf{A}^{(n)}$ is an $\alpha\times\mu_{n}$ matrix known by both GEOS 1 and GEOS 2. Under the $\bm\mu$-reconstructability
\cite{Gong2012}, the source data $\mathbf{s}$ can be reconstructed at GEOS 2 if and only if the following two conditions are met:
\begin{subequations}
\begin{align}
\mathrm{rank}&\left\{\left[\mathbf{H}^{(n)}\mathbf{A}^{(n)},n\in\mathcal{N}\right]^{T}\right\}=M,\label{equ:Rankcondition_LEO-GEO_a}\\
 &~~\mathrm{and}~\sum\limits_{n\in\mathcal{N}}\mu_{n}\geq M,\label{equ:Rankcondition_LEO-GEO_b}
 \end{align}
\end{subequations}
where $\mathrm{rank}\{\cdot\}$ denotes the rank of a matrix and $[\mathbf{H}^{(n)}\mathbf{A}^{(n)},n\in\mathcal{N}]$ denotes the matrix obtained by horizontally stacking matrices $\mathbf{H}^{(n)}\mathbf{A}^{(n)}$ for all $n\in\mathcal{N}$.
\end{itemize}
\par Note that we assume that the matrices $\left\{\mathbf{H}^{(n)},\mathbf{A}^{(n)},n\in\mathcal{N}\right\}$
are predesigned and satisfy condition (\ref{equ:Rankcondition_LEO-GEO_a}). In this work, we consider the original data reconstruction based on the $\bm\mu$-reconstructability to save energy and time in data transmission for inter-GEOS communications. 
\subsection{Communication Link Model}
\par For each GEOS, the multibeam transmitter \cite{Choi2009,Manshadi1991} is employed such that it can use the same frequency band to transmit data to multiple LEOSs simultaneously, and the multi-channel receiver is assumed such that they can receive data from different bands simultaneously. For the LEOSs, they use a common frequency band to receive data from GEOS 1 and use different bands to transmit data to GEOS 2. According to the CoRaSat scenarios defined in \cite{D2.32012,Liolis2013}, Ku band and Ka band are used for the GEOS-LEOS downlink and the LEOS-GEOS uplink, respectively.
  \par  For both types of data-links, we assume that the data transmission duration is $T$ and $t_{\mathrm{s}}$ denotes the starting time that GEOS $1/2$ begins to transmit/receive data. Thus the transmission time interval is $[t_{\mathrm{s}}, t_{\mathrm{s}}+T]$. For $n\in\mathcal{N}$, define $t_{\mathrm{s},n}$ and $t_{\mathrm{e},n}$ as the starting and ending times for transmission with LEOS $n$, respectively. Then the transmission time interval between LEOS $n$ and GEOS $1/2$ is denoted as
\begin{equation}\label{equ:conditionforTimeperiod_LEOSn}
  \mathcal{T}_{n}(t_{\mathrm{s}})=[t_{\mathrm{s},n},t_{\mathrm{e},n}]\subseteq[t_{\mathrm{s}}, t_{\mathrm{s}}+T],
\end{equation}
with $\bigcup_{n=1}^{N}\mathcal{T}_{n}(t_{\mathrm{s}})=[t_{\mathrm{s}}, t_{\mathrm{s}}+T]$.
\subsubsection{GEOS-LEOS Downlink}
\par  Let $P_{n}(t)$, $t\in[t_{\mathrm{s}}, t_{\mathrm{s}}+T]$, denote the power allocated to transmit data to LEOS $n$ with $P_{n}(t)\geq0$ for any $ t\in\mathcal{T}_{n}(t_{\mathrm{s}})$ and $P_{n}(t)=0$ otherwise. Assume that the beam of GEOS 1 to each LEOS can always be tracked in the coverage area and the receive antennas of different LEOSs can always point to GEOS 1 during the movements. Then according to \cite{Chen2014}, the channel gain of the link from GEOS $1$ to LEOS $n$ can be written as $g_{n}(t)=\frac{G_{\mathrm{T}}G_{\mathrm{R}}c ^{2}10^{(\frac{-A_{n}}{10})}}{(4\pi d_{n}(t)f)^{2}}$, where $G_{\mathrm{T}}$ and $G_{\mathrm{R}}$ denote the antenna gains (AGs) of the transceiver, $c$ denotes the velocity of light, $f$ denotes the carrier frequency, $d_{n}(t)$ is given by (\ref{equ:GEO-LEOlinkDistance}) and $A_{n}$ (measured in dB) denotes the attenuation coefficient of signal propagation. Thus the received signal-to-noise ratio (SNR) at LEOS $n$ and time $t$ is given by
\begin{equation}\label{equ:SNR_GEO-LEO}
\begin{split}
  \Gamma_{n}(t)=\frac{P_{n}(t) g_{n}(t)}{N_{0}W}&= \frac{P_{n}(t) L_{n}(f)}{d_{n}^{2}(t)},
\end{split}
\end{equation}
where $N_{0}$ denotes the noise power spectral density, $W$ denotes the assigned channel bandwidth and $L_{n}(f)\triangleq \frac{G_{\mathrm{T}}G_{\mathrm{R}}c ^{2}10^{(\frac{-A_{n}}{10})}}{(4\pi f)^{2}N_{0}W}$. Then under a capacity achieving scheme the transmission rate is
 \begin{equation}\label{equ:Capacityofdownload_function}
   R(\Gamma_{n}(t))=W\log_{2}\left(1+ \Gamma_{n}(t)\right).
 \end{equation}
  \par Since each LEOS should at least download and store $\alpha$ files (each $u$ bits) from GEOS 1 over the link time interval $[t_{\mathrm{s}},t_{\mathrm{s}}+T]$, we have
  \begin{equation}\label{equ:Capacityofdownload_GEO-LEO}
    \int_{\mathcal{T}_{n}(t_{\mathrm{s}})}R(\Gamma_{n}(\tau))\mathrm{d}\tau\geq\alpha u,~\forall n\in\mathcal{N}.
 \end{equation}
\subsubsection{LEOS-GEOS Uplink}
Suppose that the transmission bandwidths of the $N$ LEOSs are the same and again denoted by $W$. As we have assumed that each LEOS transmits data via a different frequency band, let $f_{n}$ denote the carrier frequency of LEOS $n$. Define the transmission power of LEOS $n$ at time $t$ as $P_{n}(t)$ with $P_{n}(t)\geq0$ for any $ t\in\mathcal{T}_{n}(t_{\mathrm{s}})$ and $P_{n}(t)=0$ otherwise. Similarly to (\ref{equ:SNR_GEO-LEO}), the received SNR of GEOS 2 from LEOS $n$ at time $t$ is given by
\begin{equation}\label{equ:SNRofDC_LEO-GEO}
  \Gamma_{n}(t)= \frac{P_{n}(t) L_{n}(f_{n})}{d_{n}^{2}(t)}.
\end{equation}
 \par To guarantee the $\bm\mu$-reconstructability over the LEOS-GEOS uplink during the link time interval $[t_{\mathrm{s}},t_{\mathrm{s}}+T]$, the downloaded data from LEOS $n$ should satisfy
 \begin{equation}\label{equ:RealNumFileofdownload}
  \int_{\mathcal{T}_{n}(t_{\mathrm{s}})}R( \Gamma_{n}(\tau))\mathrm{d}\tau\geq u\mu_{n},
 \end{equation}
 where $R(\cdot)$ is defined in (\ref{equ:Capacityofdownload_function}) and $\sum\limits_{n\in\mathcal{N}}\mu_n \geq M$ according to (\ref{equ:Rankcondition_LEO-GEO_b}).
\subsection{Problem Formulations}\label{sec:ProbForm:subsec:Formulations}
We now formulate problems to be solved for both the GEOS-LEOS downlink and the LEOS-GEOS uplink. 
 \subsubsection{Resource Allocation for GEOS-LEOS Downlink under Distributed-storage Coding}
 Since each LEOS downloads at least $\alpha$ coded files from GEOS 1, the objective is to minimize the total transmission energy consumption over $[t_{\mathrm{s}},t_{\mathrm{s}}+T]$ for the GEOS-LEOS downlink under the constraint that each LEOS can receive at least $\alpha$ files. We formulate the following power allocation problem:
\begin{equation}\label{equ:Powerminimization_GEO-LEO}
\begin{array}{cl}
  \displaystyle\min\limits_{\left\{P_{n}(t)\right\}_{n\in\mathcal{N}}} & \displaystyle \sum\limits_{n\in\mathcal{N}}\int_{\mathcal{T}_{n}(t_{\mathrm{s}})} {P_{n}(\tau)} {\rm{d}}\tau \\
 \displaystyle \mathrm{s.t.} &\displaystyle (\ref{equ:Capacityofdownload_GEO-LEO}),~ 0\leq P_{n}(t)\leq P_{\max},~\forall t\in\mathcal{T}_{n}(t_{\mathrm{s}}),~\forall n\in\mathcal{N},
\end{array}
\end{equation}
where the traveling-wave tube amplifier (TWTA) equipped on the satellites is considered and thus $P_{\max}$ is the maximum transmission power budget of each beam of GEOS 1 \cite{Choi2009}. Note that in this work, since higher-layer power-storage allocation is considered, we consider the compound linearized channel model incorporating the power amplifiers. And the linearization for the nonlinear power amplifiers is ignored here and one can refer to \cite{Ishiguro2013} and in the references therein for more details.
\par Another problem of interest is to minimize the transmission time given the maximum total transmission energy of GEOS 1. It is equivalent to minimizing the total GEOS-LEOS downlink time period $T$ given the starting time $t_{\mathrm{s}}$ and maximum transmission energy $E_{\max}$ and beam transmission power $P_{\max}$. Then we can formulate the transmission time minimization problem as follows,
\begin{equation}\label{equ:Timeminimization_GEO-LEO}
\begin{array}{cl}
  \displaystyle\min\limits_{\{P_{n}(t)\}_{n\in\mathcal{N}},T} &\displaystyle T\\
      \displaystyle\mathrm{s.t.}&\displaystyle\int_{t_{\mathrm{s},n}}^{t_{\mathrm{s}}+T}R(\Gamma_{n}(\tau))\mathrm{d}\tau\geq\alpha u,\\
      &\displaystyle 0\leq P_{n}(t)\leq P_{\max},~\forall t\in[t_{\mathrm{s},n},t_{\mathrm{s}}+T],~\forall n\in\mathcal{N},\\
          & \displaystyle{\sum_{n\in\mathcal{N}}\int_{t_{\mathrm{s},n}}^{t_{\mathrm{s}}+T} {P_{n}( \tau){\rm{d}}\tau } }\leq E_{\max},
\end{array}
\end{equation}
where $t_{\mathrm{s},n}$ denotes the link starting time of LEOS $n$.
\subsubsection{Resource Allocation for LEOS-GEOS Uplink under $\boldsymbol{\mu}$-Reconstructability}
\par For the data transmission and reconstruction from the $N$ LEOSs to GEOS 2 over $[t_{\mathrm{s}},t_{\mathrm{s}}+T]$, we formulate the following joint uploaded-data size and power allocation problem:
\begin{equation}\label{equ:Powerminimization_Reconst}
 \begin{array}{cl}
    \displaystyle\min\limits_{\{P_{n}(t)\}_{n\in\mathcal{N}},\bm\mu} &\displaystyle\sum\limits_{n \in\mathcal{N}} {\int_{\mathcal{T}_n(t_{\mathrm{s}})} P_{n}(\tau){\rm{d}}\tau}\\
     \mathrm{s.t.}  & \displaystyle\sum\limits_{n\in\mathcal{N}}\mu_n = M,\\
    &\begin{aligned}
      & (\ref{equ:RealNumFileofdownload}),~\mu_{n}\in\{0,1,\cdots,\alpha\},\\
      & 0\leq P_{n}(t)\leq P_{\max},~\forall t\in \mathcal{T}_n\left( {t_{\mathrm{s}}} \right),~\forall n\in\mathcal{N},
      \end{aligned}
\end{array}
\end{equation}
where $P_{\max}$ is the maximum transmission power budget of each LEOS and $\mu_{n}$ files are downloaded among the $\alpha$ files of LEOS $n$.
\par Similarly to (\ref{equ:Timeminimization_GEO-LEO}), we can also formulate the transmission time minimization problem for the LEOS-GEOS uplink as follows,
\begin{equation}\label{equ:Timeminimization_LEO-GEO}
\begin{array}{cl}
  \displaystyle\min\limits_{\{P_{n}(t)\}_{n\in\mathcal{N}},\bm\mu,T} &  T\\
      \displaystyle\mathrm{s.t.}& \displaystyle\sum\limits_{n\in \mathcal{N}} \mu_{n}=M,~\mu_{n}\in\{0,1,\cdots,\alpha\},\\
      &\displaystyle\int_{t_{\mathrm{s},n}}^{t_{\mathrm{s}}+T}R(\Gamma_{n}(\tau))\mathrm{d}\tau\geq\mu_{n}u,\\
      &\displaystyle 0\leq P_{n}(t)\leq P_{\max},~\forall t\in[t_{\mathrm{s},n},t_{\mathrm{s}}+T],~\forall n\in\mathcal{N},\\
          & \displaystyle{\sum_{n\in\mathcal{N}}\int_{t_{\mathrm{s},n}}^{t_{\mathrm{s}}+T} {P_{n}( \tau){\rm{d}}\tau } }\leq E_{\max},
\end{array}
\end{equation}
where $E_{\max}$ denotes the transmission energy budget of the whole LEOS network.
\section{GEOS-LEOS Downlink Resource Allocation}\label{sec:Solution_GEO-LEO}
\subsection{Solution to Problem (\ref{equ:Powerminimization_GEO-LEO})}\label{sec:subsubsec:Powerminimization_GEO-LEO}
Since $\{\mathcal{T}_n\left( {t_{\mathrm{s}}} \right)\}_{n\in\mathcal{N}}$ are known fixed parameters and there is no coupled constraint on $\{P_{n}(t)\}_{n\in\mathcal{N}}$, problem (\ref{equ:Powerminimization_GEO-LEO}) can be decoupled as $N$ power minimization sub-problems where the $n^{\mathrm{th}}$ sub-problem is given by
\begin{equation}\label{equ:Powerminimization_GEO-LEO1}
  \begin{array}{cl}
  \displaystyle\min\limits_{P_{n}(t)}&\displaystyle\int_{\mathcal{T}_{n}(t_{\mathrm{s}})} {P_{n}\left( \tau  \right)} {\rm{d}}\tau \\
   \displaystyle\mathrm{s.t.}&\displaystyle\int_{\mathcal{T}_{n}(t_{\mathrm{s}})}\log_{2}\left(1+ \frac{P_{n}(\tau) L_{n}(f)}{d_{n}^{2}(\tau)}\right)\mathrm{d}\tau\geq\frac{u\alpha}{W},\\
   & 0\leq P_{n}(t)\leq P_{\max},~\forall t\in \mathcal{T}_n\left( {t_{\mathrm{s}}} \right).
  \end{array}
\end{equation}
\par It is easy to see that problem (\ref{equ:Powerminimization_GEO-LEO1}) is convex. As \cite{Zhang2008} has shown that the Lagrangian coefficients of the linear power constraints in a power optimization problem can be ignored, we can directly define the following Lagrangian function
 \begin{equation}\label{equ:Lagrangefunction}
L(P_{n}( t),\lambda)=\int_{\mathcal{T}_{n}(t_{\mathrm{s}})} {P_{n}(\tau)} {\rm{d}}\tau-\lambda\left(\int_{\mathcal{T}_{n}(t_{\mathrm{s}})}\vphantom{\int_{\mathcal{T}_{n}(t_{\mathrm{s}})}\log_{2}\left(1+ \frac{P_{n}(\tau)\cdot L_{n}(f)}{d_{n}^{2}(\tau)}\right)\mathrm{d}\tau}\log_{2}\left(1+ \frac{P_{n}(\tau) L_{n}(f)}{d_{n}^{2}(\tau)}\right)\mathrm{d}\tau-\frac{u\alpha}{W}\vphantom{\int_{\mathcal{T}_{n}(t_{\mathrm{s}})}\log_{2}\left(1+ \frac{P_{n}(\tau)\cdot L_{n}(f)}{d_{n}^{2}(\tau)}\right)\mathrm{d}\tau}\right),~\forall t\in\mathcal{T}_{n}(t_{\mathrm{s}}),
\end{equation}
where $\lambda\geq0$ and $0\leq P_{n}(t)\leq P_{\max}$. The Karush-Kuhn-Tucker (KKT) conditions can then be written as
\begin{subequations}
\begin{align}
  &1+ \frac{P_{n}(t)\cdot L_{n}(f)}{d_{n}^{2}(t)}=\frac{\lambda L_{n}(f)}{d_{n}^{2}(t)\ln2},~\forall t\in\mathcal{T}_{n}(t_{\mathrm{s}}),\label{equ:Optimalcondition_a}\\
&\mathrm{and}~\int_{\mathcal{T}_{n}(t_{\mathrm{s}})}\log_{2}\left(1+ \frac{P_{n}(\tau) L_{n}(f)}{d_{n}^{2}(\tau)}\right)\mathrm{d}\tau=\frac{u\alpha}{W}.\label{equ:Optimalcondition_b}
 \end{align}
\end{subequations}
\par Taking into account $0\leq P_{n}(t)\leq P_{\max}$ and solving for $P_{n}(t)$ from (\ref{equ:Optimalcondition_a}), we obtain
\begin{equation}\label{equ:Optimalsolutionpoits_CWF}
\begin{split}
P_{n}\left( t \right) =\left[\left[{\frac{\lambda}{{\ln 2}} - \frac{{d_n^2( t)}}{{L_{n}(f)}}}\right]^ {+}\right]_{P_{\max}} ,~\forall t\in\mathcal{T}_{n}(t_{\mathrm{s}}),
\end{split}
\end{equation}
where $[x]^{+}=\max\{x,0\}$, $[x]_{a}=\min\{x,a\}$ and $\lambda$ is chosen to meet (\ref{equ:Optimalcondition_b}). To obtain the solution to $P_{n}\left( t \right)$, the time-domain constrained waterfilling algorithm is presented in \emph{Algorithm} \ref{alg:WF-PA_GEO-LEO}.
\begin{algorithm}[ht!]
\caption{Constrained waterfilling algorithm for computing (\ref{equ:Optimalsolutionpoits_CWF})}\label{alg:WF-PA_GEO-LEO}
\hspace*{0.02in} {\bf Input:} $\mathcal{T}_{n}(t_{\mathrm{s}})$, $L_{n}(f)$, $\{d_{n}(t), \forall t\in \mathcal{T}_{n}(t_{\mathrm{s}})\}$, $\alpha$, $u$, $W$, $P_{\max}$  \\
\hspace*{0.02in} {\bf Output:} $\{P_{n}(t), \forall t\in \mathcal{T}_{n}(t_{\mathrm{s}})\}$
\begin{algorithmic}[1]
\State \textbf{Initialization:} Denote $\mathcal{T}_{n,1}$ and $\mathcal{T}_{n,2}$ as the time intervals such that $P_{n}(t)=0$ and $P_{n}(t)=P_{\max}$, respectively. According to (\ref{equ:Optimalsolutionpoits_CWF}), reformulate $P_{n}(t)$ as
\[(\mathbf{A1}):P_{n}(t)=\begin{cases}\frac{\lambda}{{\ln 2}} - \frac{{d_n^2(t)}}{{L_{n}(f)}},~\forall t\in\mathcal{T}_{n}(t_{\mathrm{s}})\setminus\mathcal{T}_{n,1}\setminus\mathcal{T}_{n,2},\\
         P_{\max},~~~~~~~~~\forall t\in\mathcal{T}_{n,2},\\
         0,~~~~~~~~~~~~~\forall t\in\mathcal{T}_{n,1}. \end{cases}\]
Set $\mathcal{T}_{n,1}=\mathcal{T}_{n,2}=\emptyset$ and substitute ($\mathbf{A1}$) into (\ref{equ:Optimalcondition_b}) to obtain $\lambda= \frac{{\ln 2}}{{{L_n}\left( f \right)}}\exp \left( {\frac{{u\alpha\ln 2}}{{|\mathcal{T}_{n}(t_{\mathrm{s}})|W}}} \right)\exp \left(\frac{\int_{\mathcal{T}_{n}(t_{\mathrm{s}})} {\ln \left(d_n^2\left( t \right)\right)} \mathrm{d}t}{|\mathcal{T}_{n}(t_{\mathrm{s}})|}\right)$, where $\left|\mathcal{T}_{n}(t_{\mathrm{s}})\right|$ denotes the length of $\mathcal{T}_{n}(t_{\mathrm{s}})$. Then update $P_{n}(t)$ for all $t\in\mathcal{T}_{n}(t_{\mathrm{s}})$ by ($\mathbf{A1}$).
\State {\textbf{Update:}}
\begin{itemize}
  \item[]\hspace*{-0.28in}\textbf{while} $\exists~t$ such that $P_{n}(t)<0$ \textbf{or} $P_{n}(t)>P_{\max}$ \textbf{do}
  \item[\textbf{a.}]{\begin{itemize}
  \item \textbf{while} $\exists~t$ such that $P_{n}(t)<0$ \textbf{do}
      \begin{enumerate}
  \item Update $\mathcal{T}_{n,1}$ by $\mathcal{T}_{n,1}=\{t:P_{n}(t)\leq0\}$;
  \item Solve $\lambda$ by substituting ($\mathbf{A1}$) into (\ref{equ:Optimalcondition_b});
  \item Update $P_{n}(t)$ for all $t\in\mathcal{T}_{n}(t_{\mathrm{s}})$ by ($\mathbf{A1}$);
\end{enumerate}
 \item \textbf{end while}
\end{itemize}}
\item[\textbf{b.}]\begin{itemize}
\item \textbf{if} $\exists~t$ such that $P_{n}(t)>P_{\max}$ \textbf{do} 
\begin{enumerate}
    \item Update $\mathcal{T}_{n,2}$ by $\mathcal{T}_{n,2}=\{t:P_{n}(t)\geq P_{\max}\}$;
    \item Solve $\lambda$ by substituting ($\mathbf{A1}$) into (\ref{equ:Optimalcondition_b});
    \item Update $P_{n}(t)$ for all $t\in\mathcal{T}_{n}(t_{\mathrm{s}})$ by ($\mathbf{A1}$);
\end{enumerate}
\item \textbf{end if}
\end{itemize}
\item[] \hspace*{-0.28in}\textbf{end while}
\end{itemize}
\State \Return $\{P_{n}(t), \forall t\in \mathcal{T}_{n}(t_{\mathrm{s}})\}$.
\end{algorithmic}
\end{algorithm}
\subsection{Solution to Problem (\ref{equ:Timeminimization_GEO-LEO})}
We first prove by contradiction that the minimum transmission time is achieved by the optimal power allocation $\{P_{n}(t)\}_{n\in \mathcal{N}}$ that minimizes the total transmission energy over the given transmission interval. Assume that the optimal solution pair to problem (\ref{equ:Timeminimization_GEO-LEO}) is $(\{P_{n}^{(\mathrm{Opt})}(t)\}_{n\in\mathcal{N}},T^{(\mathrm{Opt})})$ and the transmission power minimizing the total transmission energy over $T^{(\mathrm{Opt})}$ is $\{P_{n}(t,T^{(\mathrm{Opt})})\}_{n\in\mathcal{N}}$. If $P_{n}^{(\mathrm{Opt})}(t)\neq P_{n}(t,T^{(\mathrm{Opt})})$ for certain $n\in\mathcal{N}$, then ${\sum\limits_{n\in\mathcal{N}}\int_{t_{\mathrm{s},n}}^{t_{\mathrm{s}}+T^{(\mathrm{Opt})}} {P_{n}( \tau,T^{(\mathrm{Opt})}){\rm{d}}\tau } }<\sum\limits_{n\in\mathcal{N}}\int_{t_{\mathrm{s},n}}^{t_{\mathrm{s}}+T^{(\mathrm{Opt})}} $ ${P_{n}^{(\mathrm{Opt})}( \tau){\rm{d}}\tau } $. Note that given any feasible $T$, we have that
\begin{equation}\label{equ:Timeminimization_GEO-LEO_Ineuqality}
\begin{split}
&\begin{array}{rl}
  \displaystyle \sum\limits_{n\in\mathcal{N}}\int_{t_{\mathrm{s},n}}^{t_{\mathrm{s}}+T} P_{n}( \tau,T){\rm{d}}\tau=\min\limits_{\left\{P_{n}(t)\right\}_{n\in\mathcal{N}}} & \displaystyle {\sum\limits_{n\in\mathcal{N}}\int_{t_{\mathrm{s},n}}^{t_{\mathrm{s}}+T} {P_{n}( \tau){\rm{d}}\tau } } \\
                            ~ &\displaystyle\int_{t_{\mathrm{s},n}}^{t_{\mathrm{s}}+T}\log_{2}\left(1+ \frac{P_{n}(\tau) L_{n}(f)}{d_{n}^{2}(\tau)}\right)\mathrm{d}\tau\geq\frac{u\alpha}{W},\\
                            ~ &0\leq P_{n}(t)\leq P_{\max},~\forall n\in\mathcal{N}.  \end{array}\\
 \end{split}
\end{equation}
It's easy to see that $\sum\limits_{n\in\mathcal{N}}\int_{t_{\mathrm{s},n}}^{t_{\mathrm{s}}+T} P_{n}( \tau,T){\rm{d}}\tau$ is decreasing with $T$.
Then there should exist a smaller feasible $T_{\star}$ such that $T_{\star}<T^{(\mathrm{Opt})}$ and ${\sum\limits_{n\in\mathcal{N}}\int_{t_{\mathrm{s},n}}^{t_{\mathrm{s}}+T_{\star}} {P_{n}( \tau,T_{\star}){\rm{d}}\tau } }$ $={\sum\limits_{n\in\mathcal{N}}\int_{t_{\mathrm{s},n}}^{t_{\mathrm{s}}+T^{(\mathrm{Opt})}} {P_{n}^{(\mathrm{Opt})}( \tau){\rm{d}}\tau } }$ hold. This contradicts the assumption that $T^{(\mathrm{Opt})}$ is the optimal solution. Therefore, based on the solving process of (\ref{equ:Optimalsolutionpoits_CWF}), problem (\ref{equ:Timeminimization_GEO-LEO}) can be simplified as%
\begin{equation}\label{equ:Timeminimization_GEO-LEO_Eq2}
\begin{array}{cl}
  \displaystyle\min\limits_{T} &\displaystyle T\\
      \displaystyle\mathrm{s.t.}&\displaystyle{\sum_{n\in\mathcal{N}}\int_{t_{\mathrm{s},n}}^{t_{\mathrm{s}}+T} P_{n}(\tau,T)\mathrm{d}\tau \leq E_{\max}},
\end{array}
\end{equation}
where
\begin{equation}\label{equ:Timeminimization_GEO-LEO_Eq2_sub}
P_{n}(t,T)=\left[\left[{\frac{\lambda_{n}}{{\ln 2}} - \frac{{d_n^2(t)}}{{L_{n}(f)}}}\right]^ {+}\right]_{P_{\max}},~\forall t\in[t_{\mathrm{s},n},t_{\mathrm{s}}+T],~\forall n\in\mathcal{N},
\end{equation}
and $\lambda_{n}$ is chosen such that $\int_{t_{\mathrm{s},n}}^{t_{\mathrm{s}}+T}\log_{2}\left(1+ \frac{P_{n}(\tau,T) L_{n}(f)}{d_{n}^{2}(\tau)}\right)\mathrm{d}\tau=\frac{\alpha u}{W}$.
\par It is of interest to investigate the maximum energy consumption, which corresponds to the minimum transmission duration and is characterized by $P_{n}(t,T)=P_{\max}$ during the transmission. Let $T_{n0}$ denote the minimum transmission duration with respect to LEOS $n$, then $T_{n0}$ can be obtained by using the bisection method to solve the following equations:
  \begin{equation}\label{equ:Timeminimization_GEO-LEO_critialpoint}
\begin{cases}
\displaystyle P_{n}(t,T_{n0})=P_{\max},~\forall t\in[t_{\mathrm{s},n},t_{\mathrm{s}}+T_{n0}],\\
\displaystyle \int_{t_{\mathrm{s},n}}^{t_{\mathrm{s}}+T_{n0}}\log_{2}\left(1+ \frac{P_{n}(\tau,T_{n0}) L_{n}(f)}{d_{n}^{2}(\tau)}\right)\mathrm{d}\tau=\frac{\alpha u}{W}.
\end{cases}
\end{equation}
\par Define $T_{0}\triangleq \max\limits_{n\in \mathcal{N}} T_{n0}$ and $E_{0}\triangleq\sum\limits_{n\in\mathcal{N}}\int_{t_{\mathrm{s},n}}^{t_{\mathrm{s}}+T_{0}} P_{n}(\tau,T_{0})\mathrm{d}\tau$, then $T_{0}$ denotes the common minimum transmission time for problem (\ref{equ:Timeminimization_GEO-LEO_Eq2}) and $E_{0}$ denotes the corresponding maximum transmission energy consumption when $E_{\max}\geq E_{0}$. While when $E_{\max}<E_{0}$, the optimal $T$ to problem (\ref{equ:Timeminimization_GEO-LEO_Eq2}) should achieve $\sum\limits_{n\in\mathcal{N}}\int_{t_{\mathrm{s},n}}^{t_{\mathrm{s}}+T} P_{n}(\tau,T)\mathrm{d}\tau = E_{\max}$. Thus problem (\ref{equ:Timeminimization_GEO-LEO_Eq2}) is equivalent to finding a time interval $T$ such that the total transmission energy over $T$ is equal to $E_{\max}$. Then the optimal solution to problem (\ref{equ:Timeminimization_GEO-LEO_Eq2}) can be solved by
\begin{equation}\label{equ:Timeminimization_GEO-LEO_Eq3}
\begin{cases}
{\displaystyle\sum\limits_{n\in\mathcal{N}}\int_{t_{\mathrm{s},n}}^{t_{\mathrm{s}}+T} P_{n}(\tau,T)\mathrm{d}\tau = E_{\max}},\\
\displaystyle P_{n}(t,T)=\left[\left[{\frac{\lambda_{n}}{{\ln 2}} - \frac{{d_n^2(t)}}{{L_{n}(f)}}}\right]^ {+}\right]_{P_{\max}},\forall t\in[t_{\mathrm{s},n},t_{\mathrm{s}}+T],\\
\displaystyle \int_{t_{\mathrm{s},n}}^{t_{\mathrm{s}}+T}\log_{2}\left(1+ \frac{P_{n}(\tau,T) L_{n}(f)}{d_{n}^{2}(\tau)}\right)\mathrm{d}\tau=\frac{\alpha u}{W},~\forall n\in\mathcal{N}.
\end{cases}
\end{equation}
\par As given $T$, $P_{n}(t,T)$ for all $n\in \mathcal{N}$ can be obtained by \emph{Algorithm} \ref{alg:WF-PA_GEO-LEO}, and thus the corresponding total transmission energy can be obtained. Based on the decreasing property of $\sum\limits_{n\in\mathcal{N}}\int_{t_{\mathrm{s},n}}^{t_{\mathrm{s}}+T} P_{n}(t,T)\mathrm{d}t$ with respect to $T$ according to (\ref{equ:Timeminimization_GEO-LEO_Ineuqality}), we can use the bisection method to obtain $T$ that satisfies (\ref{equ:Timeminimization_GEO-LEO_Eq3}). The detailed procedure is presented in \emph{Algorithm} \ref{alg:GEO-LEO_MiniTime}.
\begin{algorithm}
\caption{Bisection-based constrained waterfilling algorithm for solving problem (\ref{equ:Timeminimization_GEO-LEO})}\label{alg:GEO-LEO_MiniTime}
\hspace*{0.02in} {\bf Input:} $N$, $M$, $\alpha$, $u$, $W$, $\{L_{n}(f),t_{\mathrm{s},n}\}_{n\in\mathcal{N}}$, $t_{\mathrm{s}}$, $P_{\max}$, $E_{\max}$\\
\hspace*{0.02in} {\bf Output:} $T$, $\{P_{n}(t), \forall t\in[t_{\mathrm{s},n}, t_{\mathrm{s}}+T]\}_{n\in\mathcal{N}}$
\begin{algorithmic}[1]
\State \textbf{Initialization:} Set $E=0$, $T_{\min}=0$ and assign a large feasible value to $T_{\max}$.
\State \textbf{Step 1:} Solve $\{T_{n0}\}_{n\in\mathcal{N}}$ based on (\ref{equ:Timeminimization_GEO-LEO_critialpoint}) via the bisection method since $\int_{t_{\mathrm{s},n}}^{t_{\mathrm{s}}+T_{n0}}$ $\log_{2}\left(1+ \frac{P_{\max}  L_{n}(f)}{d_{n}^{2}(\tau)}\right)\mathrm{d}\tau$ is increasing with respect to $T_{n0}$;
\State \textbf{Step 2:} Set $T_{0}\triangleq \max\limits_{n\in \mathcal{N}} T_{n0}$ and use \emph{Algorithm} \ref{alg:WF-PA_GEO-LEO} to calculate $E_{0}=\sum\limits_{n\in\mathcal{N}}\int_{t_{\mathrm{s},n}}^{t_{\mathrm{s}}+T_{0}} P_{n}(\tau,T_{0})\mathrm{d}\tau$, where $P_{n}(\tau,T)$ is given by (\ref{equ:Timeminimization_GEO-LEO_Eq2_sub});
\State \textbf{Step 3:} Solve problem (\ref{equ:Timeminimization_GEO-LEO_Eq2}) according to the relationship between $E_{\max}$ and $E_{0}$:
\begin{itemize}
  \item[]\hspace*{-0.28in}\textbf{if} $E_{\max}\geq E_{0}$ \textbf{do}
    \item[]\hspace*{-0.28in}~Set $T=T_{0}$ and $P_{n}(t)=P_{n}(t,T_{0})$ given by (\ref{equ:Timeminimization_GEO-LEO_Eq2_sub}) for $n\in\mathcal{N}$;
  \item[]\hspace*{-0.28in}\textbf{else}
    \item[]\hspace*{-0.28in}~Solve $T$ and $\{P_{n}(t)\}_{n\in \mathcal{N}}$ based on (\ref{equ:Timeminimization_GEO-LEO_Eq3}) via the bisection method using \emph{Algorithm} \ref{alg:WF-PA_GEO-LEO}:
           \item[\textbf{--}]\textbf{while} $E\neq E_{\max}$ \textbf{do}
            \begin{enumerate}
                   \item Update $T=(T_{\min}+T_{\max})/2$ and update $E=\sum\limits_{n\in\mathcal{N}}\int_{t_{\mathrm{s},n}}^{t_{\mathrm{s}}+T} P_{n}(\tau,T)\mathrm{d}\tau$ via \emph{Algorithm} \ref{alg:WF-PA_GEO-LEO};
                    \item \textbf{If} $E<E_{\max}$ \textbf{then} $T_{\max}=T$; \textbf{otherwise}, $T_{\min}=T$.
                \end{enumerate}
           \item[\textbf{--}]\textbf{end while}
            \item[]\hspace*{-0.28in}\textbf{end if}
\end{itemize}
\State \Return $T$, $\{P_{n}(t), \forall t\in[t_{\mathrm{s},n}, t_{\mathrm{s}}+T]\}_{n\in\mathcal{N}}$.
\end{algorithmic}
\end{algorithm}
\section{LEOS-GEOS Uplink Resource Allocation}\label{sec:Solution_LEO-GEO}
\subsection{Solution to Problem (\ref{equ:Powerminimization_Reconst})}\label{sec:subsec:LEO-GEO_energyMinimization}
For notational simplicity, denote $\mathfrak{p}_{n}=\{P_{n}(t),\forall t\in\mathcal{T}_{n}(t_{\mathrm{s}})\}$ and $\mathcal{P}=\{\mathfrak{p}_{n}\}_{n\in\mathcal{N}}$. Then we re-write (\ref{equ:Powerminimization_Reconst}) as
\begin{equation}\label{equ:Powerminimization_Reconst_Eq1}
 \begin{array}{cl}
    \displaystyle\min\limits_{\mathcal{P},\bm\mu} &\displaystyle f_{\mathrm{E}}(\mathcal{P})\\
     \displaystyle\mathrm{s.t.}  & \displaystyle\sum\limits_{n\in\mathcal{N}}\mu_n = M,~\mu_{n}\in\{0,1,\cdots,\alpha\},\\
    &\displaystyle\begin{aligned}
      & f_{n}(\mathfrak{p}_{n},\mu_{n})\leq 0,~\forall n\in\mathcal{N},\\
      & 0\leq \mathcal{P}\leq P_{\max},
      \end{aligned}
\end{array}
\end{equation}
where
\begin{equation}\label{equ:Powerminimization_Reconst_Eq1_sub}
  \begin{split}
    f_{\mathrm{E}}(\mathcal{P})&\triangleq\sum\limits_{n \in\mathcal{N}} {\int_{\mathcal{T}_n(t_{\mathrm{s}})} P_{n}(\tau){\rm{d}}\tau}, \\
     f_{n}(\mathfrak{p}_{n},\mu_{n})&\triangleq\frac{u\mu_{n}}{W}-\int_{\mathcal{T}_{n}(t_{\mathrm{s}})}\log_{2}\left(1+ \frac{P_{n}(\tau) L_{n}(f_{n})}{d_{n}^{2}(\tau)}\right)\mathrm{d}\tau.
  \end{split}
\end{equation}
\par Problem (\ref{equ:Powerminimization_Reconst_Eq1}) is a mixed integer nonlinear program (MINLP)\cite{Bussieck2003,Chen2017}, which will be solved using the outer approximation (OA) method \cite{Duran1986,Belotti2013}. The main idea of the OA method is to repeatedly solve a linear relaxation of the MINLP, known as the master problem, which is obtained by linearizing the nonlinear constraints at the OA-point set. The OA-point set is initialized by the solution to the continuous relaxation of the MINLP and then updated by accumulating all the solutions to the master problem that have been obtained. The feasible region is updated by the lower and upper bounds which are obtained by the solution to the master problem and the feasible solution to the MINLP, respectively. The main theoretical justification is that the OA method can achieve the same optimal solution to the MINLP if the following two main conditions hold \cite{Belotti2013}: the variable-dependent nonlinear functions in the problem are convex and twice continuously differentiable, and the feasible set of the problem is a bounded polyhedral set and contains a strictly interior feasible point. For problem (\ref{equ:Powerminimization_Reconst_Eq1}), it is easy to see that the program is convex and twice continuously differentiable and its feasible set is a bounded polyhedral. As the case at the boundary point can be solved separately, then the program should contain a strictly interior feasible point if it is solvable. Thus problem (\ref{equ:Powerminimization_Reconst_Eq1}) satisfies the above two conditions and can be solved by the OA method. The procedure for solving (\ref{equ:Powerminimization_Reconst_Eq1}) based on the OA method is derived in Appendix \ref{app:OAmethod_LEC-INLP} and summarized in \emph{Algorithm} \ref{alg:OA-DSA}.
\begin{algorithm}[ht!]
\caption{OA algorithm for solving problem (\ref{equ:Powerminimization_Reconst_Eq1})}\label{alg:OA-DSA}
\hspace*{0.02in} {\bf Input:} $N$, $M$, $\alpha$, $u$, $W$, $\{L_{n}(f_{n})\}_{n\in\mathcal{N}}$ $\{\mathcal{T}_{n}(t_{\mathrm{s}})\}_{n\in\mathcal{N}}$.\\
\hspace*{0.02in} {\bf Output:} $(\mathcal{P},\bm\mu)$.
\begin{algorithmic}[1]
\State \textbf{Initialization:} Choose a tolerance $\epsilon\geq0$, set the lower bound $z_{\mathrm{L}}=-\infty$ and the upper bound $z_{\mathrm{U}}=+\infty$ for $f_{\mathrm{E}}(\mathcal{P})$ given in (\ref{equ:Powerminimization_Reconst_Eq1_sub}) and set $k=0$ and $\mathcal{S}_{\mathcal{P},\bm\mu}=\varnothing$.
\State \textbf{Step 1:} Solve $\mathrm{NLPR}$ (the continuous relaxation of (\ref{equ:Powerminimization_Reconst_Eq1})) formulated in (\ref{equ:PowerminimizationLEO-GEO_Relaxation_NLPR}) using a convex solver and obtain its optimal solution pair $(\mathcal{P}_{\min}^{(\mathrm{NLPR})},\bm\mu_{\min}^{(\mathrm{NLPR})})$. If $\bm\mu_{\min}^{(\mathrm{NLPR})}$ is an integer vector, then $f_{\mathrm{E}}(\mathcal{P}_{\min}^{(\mathrm{NLPR})})$ is the optimal value of (\ref{equ:Powerminimization_Reconst_Eq1}), thus return $(\mathcal{P}_{\min}^{(\mathrm{NLPR})},\bm\mu_{\min}^{(\mathrm{NLPR})})$ and stop; otherwise, set $\mathcal{S}_{\mathcal{P},\bm\mu}=\{(\mathcal{P}_{\min}^{(\mathrm{NLPR})},\bm\mu_{\min}^{(\mathrm{NLPR})})\}$.
\State \textbf{Step 2:} OA:
\While{$z_{\mathrm{U}}-z_{\mathrm{L}}>\epsilon$}
\begin{itemize}
  \item [1)] Solve $\mathrm{OA}$-${\mathrm{ILP}}$ formulated in (\ref{equ:PowerminimizationLEO-GEO_OA-ILP}) using an MILP solver and obtain its optimal solution pair $(\mathcal{P}_{\mathrm{OA}}^{(k)},\bm\mu_{\mathrm{OA}}^{(k)})$;
      \item [2)] Set $\widehat{\bm\mu}:=\bm\mu_{{\mathrm{OA}}}^{(k)}$ and solve the $\mathrm{NLP}$ formulated in (\ref{equ:PowerminimizationLEO-GEO_NLP}) using a convex
solver to obtain its optimal solution $\mathcal{P}_{\mathrm{NLP}}^{(k)}$;
  \item [3)] Set $z_{\mathrm{L}}=f_{\mathrm{E}}(\mathcal{P}_{\mathrm{OA}}^{(k)})$ and $z_{\mathrm{U}}=f_{\mathrm{E}}(\mathcal{P}_{\mathrm{NLP}}^{(k)})$;
  \item [4)] Update $\mathcal{S}_{\mathcal{P},\bm\mu}:=\mathcal{S}_{\mathcal{P},\bm\mu}\bigcup\{(\mathcal{P}_{\mathrm{NLP}}^{(k)},\bm\mu_{\mathrm{OA}}^{(k)})\}$ and $k:=k+1$.
\end{itemize}
\EndWhile\State{\textbf{end while}}
\State \Return $(\mathcal{P}_{\mathrm{NLP}}^{(k)},\bm\mu_{\mathrm{OA}}^{(k)})$.
\end{algorithmic}
\end{algorithm}
\subsection{Solution to Problem (\ref{equ:Timeminimization_LEO-GEO})}
\par Note that $P_{n}(t)=P_{\max}$ characterizes both the minimum transmission duration and the maximum transmission energy consumption. Let $T_{0}$ and $E_{0}$ denote the corresponding transmission duration and energy consumption, respectively. Then according to (\ref{equ:Timeminimization_LEO-GEO}), $T_{0}$ can be obtained by
\begin{equation}\label{equ:Timeminimization_LEO-GEO_sub1}
\begin{array}{cl}
  \displaystyle T_{0}=\min\limits_{\bm\mu,T} &  T\\
      \displaystyle~~~~~~~\mathrm{s.t.}& \displaystyle\sum\limits_{n\in \mathcal{N}} \mu_{n}=M,~\mu_{n}\in\{0,1,\cdots,\alpha\},\\
      &\displaystyle\frac{W}{u}\int_{t_{\mathrm{s},n}}^{t_{\mathrm{s}}+T}\log_{2}\left(1+ \frac{P_{\max} L_{n}(f_{n})}{d_{n}^{2}(\tau)}\right)\mathrm{d}\tau\geq\mu_{n},~\forall n\in\mathcal{N},
\end{array}
\end{equation}
and then
\begin{equation}\label{equ:Timeminimization_LEO-GEO_sub2}
\begin{array}{cl}
  \displaystyle E_{0}=\min\limits_{\left\{P_{n}(t)\right\}_{n\in\mathcal{N}},\bm\mu} & \displaystyle \sum\limits_{n\in\mathcal{N}}\int_{t_{\mathrm{s},n}}^{t_{\mathrm{s}}+T_{0}} {P_{n}(\tau)} {\rm{d}}\tau,\\
      \displaystyle~~~~~~~\mathrm{s.t.}& \displaystyle\sum\limits_{n\in \mathcal{N}} \mu_{n}=M,~\mu_{n}\in\{0,1,\cdots,\alpha\},\\
      &\displaystyle\frac{W}{u}\int_{t_{\mathrm{s},n}}^{t_{\mathrm{s}}+T_{0}}\log_{2}\left(1+ \frac{P_{n}(\tau) L_{n}(f_{n})}{d_{n}^{2}(\tau)}\right)\mathrm{d}\tau\geq\mu_{n},~\forall n\in\mathcal{N}.
\end{array}
\end{equation}
\par For problem (\ref{equ:Timeminimization_LEO-GEO_sub1}), it can be solved by the bisection method, where $T$ is adjusted to guarantee that the summation $\sum\limits_{n \in {\cal N}}\mu_n = M$ via the following equation:
\begin{equation}\label{equ:integervalueoperation1}
  f(T)\triangleq\sum_{n\in\mathcal{N}}\left\lfloor\frac{W}{u}\int_{t_{\mathrm{s},n}}^{t_{\mathrm{s}}+T}\log_{2}\left(1+ \frac{P_{\max}L_{n}(f_{n})}{d_{n}^{2}(\tau)}\right)\mathrm{d}t\right\rfloor=M,
\end{equation}
where $\lfloor\cdot\rfloor$ denotes the flooring operation, and in each round of search the lower bound $T_{\min}$ and upper bound $T_{\max}$ of $T$ are updated by $f(T_{\min})<M$ and $f(T_{\max})=M$, respectively. The flooring operation is to guarantee that LEOS $n$ for $n\in \mathcal{N}$ should transmit at least integer $\mu_{n}$ files while the way of setting $T_{\min}$ and $T_{\max}$ is to find the minimal $T_{0}$ such that at least one LEOS can exactly transmit integer number of files under the maximum transmission power $P_{\max}$, i.e.,
\begin{equation}\label{equ:integervalueoperation2}
\begin{split}
  \frac{W}{u}\int_{t_{\mathrm{s},n}}^{t_{\mathrm{s}}+T_{0}}\log_{2}\left(1+ \frac{P_{\max}L_{n}(f_{n})}{d_{n}^{2}(\tau)}\right)\mathrm{d}\tau&=\left\lfloor\frac{W}{u}\int_{t_{\mathrm{s},n}}^{t_{\mathrm{s}}+T_{0}}\log_{2}\left(1+ \frac{P_{\max}L_{n}(f_{n})}{d_{n}^{2}(\tau)}\right)\mathrm{d}\tau\right\rfloor\\
  &=\mu_{n},~\mathrm{for}~\mathrm{certain}~n\in\mathcal{N}.
  \end{split}
\end{equation}
It's easy to prove by contradiction that $T_{0}$ is the optimal solution to problem (\ref{equ:Timeminimization_LEO-GEO_sub1}). While for problem (\ref{equ:Timeminimization_LEO-GEO_sub2}), it is a convex and twice continuously differentiable MINLP and is equivalent to problem (\ref{equ:Powerminimization_Reconst}), and thus the OA method proposed in Section \ref{sec:subsec:LEO-GEO_energyMinimization} can be used to solve it.
\par Based on the solutions obtained from (\ref{equ:Timeminimization_LEO-GEO_sub1}) and (\ref{equ:Timeminimization_LEO-GEO_sub2}), when $E_{\max}\geq E_{0}$, the minimum transmission time is $T_{0}$; while when $E_{\max}< E_{0}$, we can have that the optimal $T$ achieves that $\min\limits_{\{P_{n}(t)\}_{n\in\mathcal{N}},\bm\mu}\sum\limits_{n\in\mathcal{N}}\int_{t_{\mathrm{s},n}}^{t_{\mathrm{s}}+T} {P_{n}( \tau){\rm{d}}\tau }=E_{\max}$, then problem (\ref{equ:Timeminimization_LEO-GEO}) is equivalent to finding a time interval $T$ to solve the following equation:
\begin{equation}\label{equ:Timeminimization_LEO-GEO_SearchProblem}
\begin{array}{cl}
  \displaystyle E_{\max}=\min\limits_{\{P_{n}(t)\}_{n\in\mathcal{N}},\bm\mu} &  \displaystyle\sum_{n\in\mathcal{N}}\int_{t_{\mathrm{s},n}}^{t_{\mathrm{s}}+T} {P_{n}( \tau){\rm{d}}\tau } \\
      ~~~~~~~~~~\displaystyle\mathrm{s.t.}& \displaystyle\sum\limits_{n\in \mathcal{N}} \mu_{n}=M,~\mu_{n}\in\{0,1,\cdots,\alpha\},\\
      &\displaystyle\int_{t_{\mathrm{s},n}}^{t_{\mathrm{s}}+T}R(\Gamma_{n}(\tau))\mathrm{d}\tau\geq\mu_{n}u,\\
      &\displaystyle 0\leq P_{n}(t)\leq P_{\max},~\forall t\in[t_{\mathrm{s},n},t_{\mathrm{s}}+T],~\forall n\in\mathcal{N}.
\end{array}
\end{equation}
\par Note that the optimization on the right-hand side of (\ref{equ:Timeminimization_LEO-GEO_SearchProblem}) is equivalent to problem (\ref{equ:Powerminimization_Reconst}) under a given $T$ while the optimal value can be viewed as a decreasing function of $T$, thus a combination of the OA method proposed in Section \ref{sec:subsec:LEO-GEO_energyMinimization} and the bisection method can be used to solve (\ref{equ:Timeminimization_LEO-GEO_SearchProblem}). The procedure for solving problem (\ref{equ:Timeminimization_LEO-GEO}) is summarized in \emph{Algorithm} \ref{alg:LEO-GEO_MiniTime}.
\begin{algorithm}[ht!]
\caption{Bisection-based OA algorithm for solving problem (\ref{equ:Timeminimization_LEO-GEO})}\label{alg:LEO-GEO_MiniTime}
\hspace*{0.02in} {\bf Input:} $N$, $M$, $\alpha$, $u$, $W$, $\{L_{n}(f_{n}),t_{\mathrm{s},n}\}_{n\in\mathcal{N}}$, $t_{\mathrm{s}}$, $P_{\max}$, $E_{\max}$\\
\hspace*{0.02in} {\bf Output:} $T$, $\{P_{n}(t), \forall t\in[t_{\mathrm{s},n}, t_{\mathrm{s}}+T]\}_{n\in\mathcal{N}}$, $\bm\mu$
\begin{algorithmic}[1]
\State \textbf{Initialization:} Set $E=0$, $T_{\min}=0$ and assign a large feasible value to $T_{\max}$.
\State \textbf{Step 1:} Solve $T_{0}$ based on (\ref{equ:Timeminimization_LEO-GEO_sub1}) via the bisection method since $f(T)$ defined in (\ref{equ:integervalueoperation1}) is non-decreasing with respect to $T$;
\State \textbf{Step 2:} Solve $E_{0}$ based on (\ref{equ:Timeminimization_LEO-GEO_sub2}) via \emph{Algorithm} \ref{alg:OA-DSA};
\State \textbf{Step 3:} Solve problem (\ref{equ:Timeminimization_LEO-GEO}) according to the relationship between $E_{\max}$ and $E_{0}$:
\begin{itemize}
  \item[]\hspace*{-0.28in}\textbf{if} $E_{\max}\geq E_{0}$ \textbf{do}
    \item[]\hspace*{-0.28in}~Set $T=T_{0}$ and update $\{P_{n}(t), \forall t\in[t_{\mathrm{s},n}, t_{\mathrm{s}}+T]\}_{n\in\mathcal{N}}$ and $\bm\mu$ according to the solution ~~~~ \hspace*{-0.28in}~obtained from (\ref{equ:Timeminimization_LEO-GEO_sub2});
  \item[]\hspace*{-0.28in}\textbf{else}
    \item[]\hspace*{-0.28in}~Solve $T$, $\{P_{n}(t)\}_{n\in \mathcal{N}}$ and $\bm\mu$ based on (\ref{equ:Timeminimization_LEO-GEO_SearchProblem}) via the bisection method using \emph{Algorithm} \ref{alg:OA-DSA}:
            \item[\textbf{--}]\textbf{while} $E\neq E_{\max}$ \textbf{do}
            \begin{enumerate}
                   \item Update $T=(T_{\min}+T_{\max})/2$ and update $E=\sum\limits_{n\in\mathcal{N}}\int_{t_{\mathrm{s},n}}^{t_{\mathrm{s}}+T} P_{n}(\tau)\mathrm{d}\tau$ by solving (\ref{equ:Timeminimization_LEO-GEO_SearchProblem}) via \emph{Algorithm} \ref{alg:OA-DSA};
                    \item \textbf{If} $E<E_{\max}$ \textbf{then} $T_{\max}=T$; \textbf{otherwise}, $T_{\min}=T$.
           \end{enumerate}
           \item[\textbf{--}]\textbf{end while}
            \item[]\hspace*{-0.28in}\textbf{end if}
\end{itemize}
\State \Return $T$, $\{P_{n}(t), \forall t\in[t_{\mathrm{s},n}, t_{\mathrm{s}}+T]\}_{n\in\mathcal{N}}$, $\bm\mu$.
\end{algorithmic}
\end{algorithm}
\section{Resouce Allocation for Failed LEOS Repair}\label{sec:RRforFailedLEOS}
\par Since the LEOSs work as distributed storage and relay nodes, the reliability of the data stored in them is very important. If the data assigned to any LEOS are failed due to the unpredictable in-orbit problems such as link errors and memory errors, then they can be efficiently regenerated by downloading $\gamma=D\beta$ files from any other $D$ LEOSs according to Section \ref{sec:ProbForm}. Therefore, in this section we provide a preliminary discussion on the issue of failed data regeneration for any LEOS.
\par According to the conditions of the optimal regenerating codes given in (\ref{equ:MSR_GEO-LEO}) and (\ref{equ:MBR_GEO-LEO}), we have the following:
\begin{itemize}
  \item for the MSR point, the data in a failed LEOS can be regenerated by choosing any other $D=\frac{\alpha}{\beta}+K-1$ LEOSs and downloading $\beta$ files from each of them;
  \item for the MBR point, the data in a failed LEOS can be regenerated by choosing any $D=\frac{\alpha}{\beta}$ LEOSs and downloading $\beta$ files from each of them.
\end{itemize}
\par Assume that the failed LEOS is $n_{0}$ and denote $\beta_{n}$ as the number of files downloaded from LEOS $n\in\mathcal{N}\setminus\{n_{0}\}$ by LEOS $n_{0}$. The conditions for data regeneration at LEOS $n_{0}$ for the above two optimal points is
$\sum\limits_{n\in\mathcal{N}\setminus\{n_{0}\}}\beta_{n}=D$ and $\beta_{n}\in\{0,\beta\}$ for any $n\in\mathcal{N}$, based on which resource allocations for total transmission energy and time minimizations can then be formulated similarly as (\ref{equ:Powerminimization_Reconst}) and (\ref{equ:Timeminimization_LEO-GEO}), respectively. And the corresponding problems can be solved similarly as those methods presented in Section \ref{sec:Solution_LEO-GEO}.
\section{Numerical Results}\label{sec:NumResult}
\par In this section, we perform numerical simulations to evaluate the proposed algorithms. The system parameters are presented in Table \ref{tab:Params_SimRes}. Assume that there are $N=5$ LEOSs in the space and the number of source files initially stored on GEOS 1 is $M=30$ and the regenerating code at the MSR point is used. The GEOS coverage angle is $\theta_{\mathrm{G}}=12^{\mathrm{o}}$, which means the GEOS coverage area is from $-6^{\mathrm{o}}$ to $+6^{\mathrm{o}}$ as seen from the sub-satellite point of the GEOS. According to \cite{Sharma2016,Chen2014}, the path attanuation (sometimes counted in another way by fade margin) in satellite communications usually ranges from several decibels to a dozen decibels. Thus $\{A_{n}\}_{n\in \mathcal{N}}$ in this paper are uniformly chosen over $[0~10]\mathrm{dB}$ based on the transmission distance. In the LEOS-GEOS uplink, the transmit carrier frequencies of the 5 LEOSs are equi-spaced. For each type of data-links, suppose that LEOS $5$ is the first satellite that enters the GEOS coverage area and set the entering time by $t=0$. The values of the initial angle difference $\{\phi_{0n}\}_{n\in\mathcal{N}}$ presented in Table \ref{tab:Params_SimRes} denote the angle differences (in absolute values) at $t=0$ between LEOS $n$ ($\in\mathcal{N}$) and LEOS $5$ as seen from the center point of the Earth. Then the instantaneous rotation angle $\varphi_{n}(t)$ for $n\in\mathcal{N}$ is given by
  \begin{equation}\label{equ:LEOinstantaneousAngle}
  \displaystyle{\varphi _n}(t) = \frac{v_{n}t}{R_{\mathrm{L},n}} - {\phi _{0n}}+\varphi_{5}(0),~t\geq0,
  \end{equation}
  where $v_n$ denotes the velocity of LEOS $n$, $R_{\mathrm{L},n}=H_{\mathrm{L},n}+R_{\mathrm{E}}$ and $\varphi_{5}(0)=-41.06^{\mathrm{o}}$ which is obtained according to the GEOS-LEOS geometry and value of $\theta_{\mathrm{G}}$.
\begin{table}[ht!]
  \centering
  \caption{Simulation parameters}\label{tab:Params_SimRes}
\begin{tabular}{c|c}
  \hline
  \textbf{Distributed-storage coding parameters}&\textbf{value}\\
   \hline
  MSR:$(M,N,K,D,\alpha,\beta)$& $(30,5,3,4, 10,5)$\\
  File size $u$& $20$MB\\
  \hline
   \textbf{Basic space network parameters} &\textbf{value} \\
  \hline
   GEOS altitude $H_{\mathrm{G}}$ & $35786$km  \\
   Earth radius $R_{\mathrm{E}}$ & $6371$km \\
   GEOS coverage angle $\theta_{\mathrm{G}}$&$12^{\mathrm{o}}$\\
   LEOS altitudes $\{H_{\mathrm{L},n}\}_{n\in\mathcal{N}}$& $[500~700~900~1100~1300]$km \\
   LEOS velocities $\{v_{n}\}_{n\in\mathcal{N}}$& $[7.2~7.3~7.4~7.5~7.6]$km/s \\
   Initial angle difference $\{\phi_{0n}\}_{n\in\mathcal{N}}$& $[12^{\mathrm{o}}~9^{\mathrm{o}}~6^{\mathrm{o}}~3^{\mathrm{o}}~0^{\mathrm{o}}]$ \\
   Path attenuation $\{A_{n}\}_{n\in\mathcal{N}}$ & $[10~8~6~4~2]$dB \\
   \hline
   \textbf{GEOS-LEOS downlink parameters} &\textbf{value} \\
  \hline
   GEOS carrier $f$ $\&$ bandwidth $W$&$19.7\mathrm{GHz}$$\&$$40\mathrm{MHz}$\\
   AGs $ G_{\mathrm{T}}\&\{G_{\mathrm{R},n}\}_{n\in\mathcal{N}}$ &$40\mathrm{dB}\&10\mathrm{dB}$  \\
    Noise power $N_{0}$   & -126.56dB  \\
  \hline
   \textbf{LEOS-GEOS uplink parameters} &\textbf{value} \\
  \hline
   LEOS carriers $\{f_{n}\}_{n\in\mathcal{N}}$ &$29.5-31\mathrm{GHz}$ (5 carriers)\\
   LEOS bandwidth $W$&$20\mathrm{MHz}$\\
   AGs $ \{G_{\mathrm{T},n}\}_{n\in\mathcal{N}}\& G_{\mathrm{R}}$ &$20\mathrm{dB}\&20\mathrm{dB}$  \\
    Noise power $N_{0}$   & -129.08dB  \\
  \hline
\end{tabular}
\end{table}
\subsection{Results for the GEOS-LEOS Downlink}\label{sec:SimResults:subsec:GEO-LEO}
\par
\begin{figure}[ht!]
   \subfigure[]
  {\begin{minipage}{1\textwidth}
  \centering
  \includegraphics[scale=0.8]{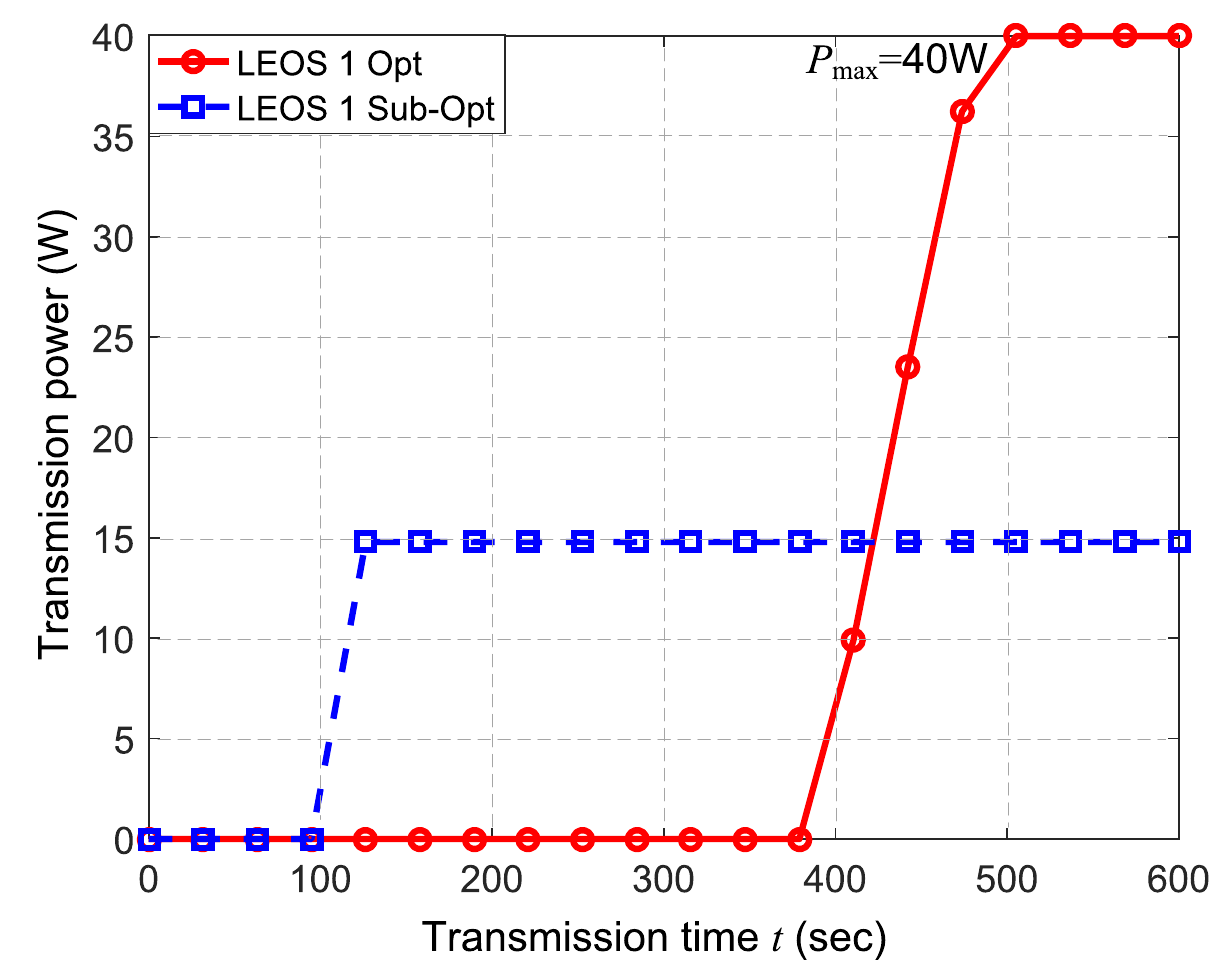}\label{fig:subfig:GEO-LEOlink_MTTE_TTP}
  \end{minipage}}\\
   \subfigure[]
  {\begin{minipage}{1\textwidth}
  \centering
  \includegraphics[scale=0.8]{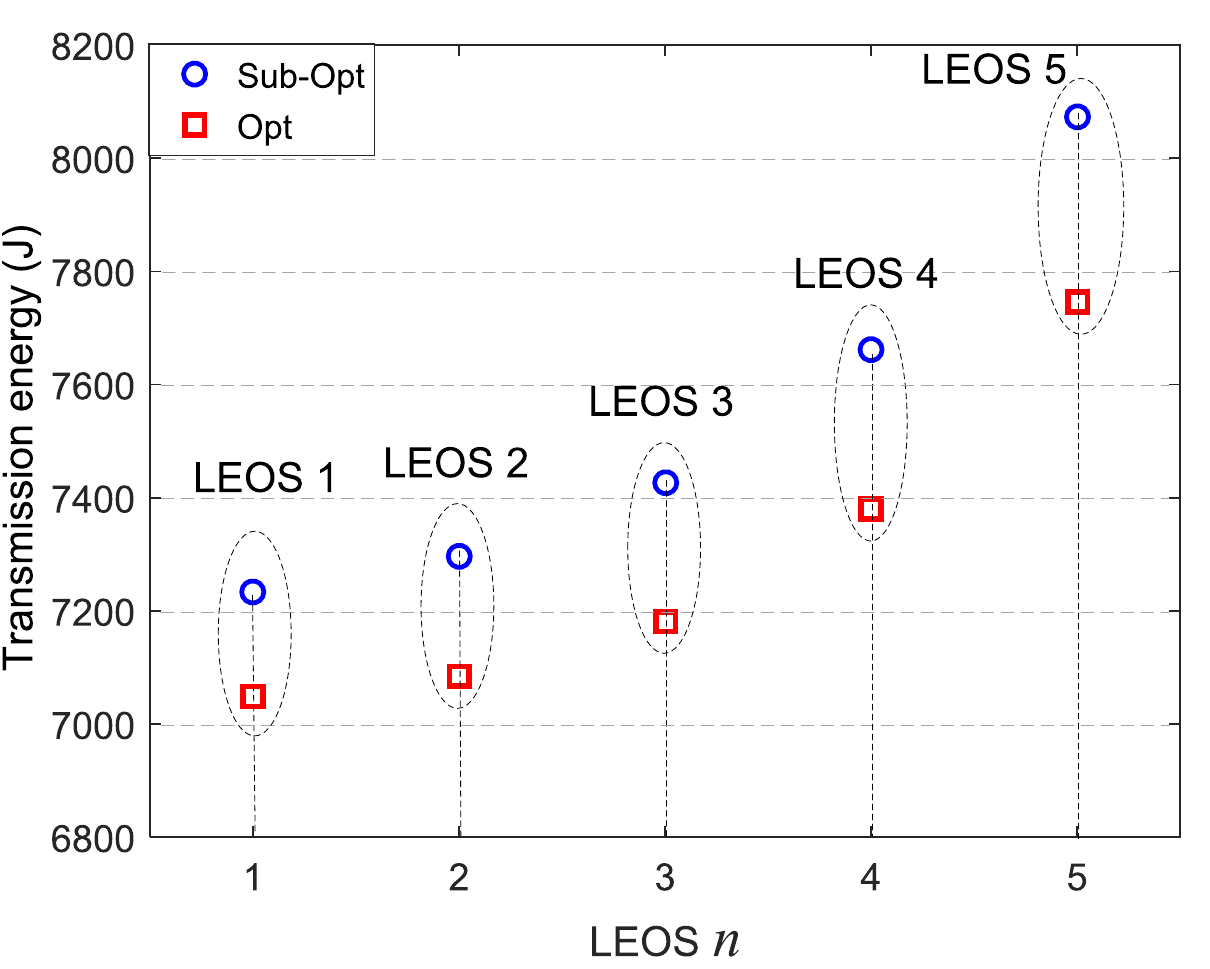}\label{fig:subfig:GEO-LEOlink_MTTE_TE}
  \end{minipage}}
  \caption{Optimization results for minimizing transmission energy over the GEOS-LEOS downlink with $T=600$s. (a) Transmission power v.s. transmission time for starting time $t_{\mathrm{s}}=0$s. (b) The transmission energy of GEOS 1 for starting time $t_{\mathrm{s}}=0$s.}\label{fig:Fig_LEO-GEOlink_AlgorithmVerification}
  \end{figure}

 We first compare the transmission power allocation algorithm proposed in Section \ref{sec:subsubsec:Powerminimization_GEO-LEO} with the sub-optimal constant power allocation strategy that satisfies the constraint:
\begin{equation}\label{equ:CPA_GEO-LEO}
 \begin{cases}\displaystyle\int_{\mathcal{T}_{n}(t_{\mathrm{s}})}R\left(\frac{P_{n}\cdot L_{n}(f)}{d_{n}^{2}(\tau)}\right)\mathrm{d}\tau=\alpha u,\\
 \displaystyle0\leq P_{n}\leq P_{\max},~\forall n\in \mathcal{N},
 \end{cases}
\end{equation}
where $\mathcal{T}_{n}(t_{\mathrm{s}})$ is defined in (\ref{equ:conditionforTimeperiod_LEOSn}), $R(\cdot)$ is defined in (\ref{equ:Capacityofdownload_function}), $P_{n}$ denotes the constant transmission power allocated to transmit data to LEOS $n$, $d_{n}^{2}(\tau)$ and $L_{n}(f)$ are defined in (\ref{equ:GEO-LEOlinkDistance}) and (\ref{equ:SNR_GEO-LEO}), respectively. Then the bisection method can be used to solve $P_{n}$ for $n\in\mathcal{N}$ based on (\ref{equ:CPA_GEO-LEO}). Set the maximum transmission power constraint for each beam of GEOS 1 by $P_{\max}=40$W. Set the total transmission time period $T=600$s, and for all $n\in\mathcal{N}$, let $t_{\mathrm{s},n}=\max\{t_{\mathrm{s}},t_{0n}\}$ and $t_{\mathrm{e},n}=t_{\mathrm{s}}+T$, which denote the starting and ending times of GEOS 1 transmitting data to LEOS $n$, respectively, with $t_{\mathrm{s}}$ representing the starting transmission time of GEOS 1 for the LEOS network and $t_{0n}$ representing the enter time of LEOS $n$ to the coverage area of GEOS 1 and being calculated by (\ref{equ:LEOinstantaneousAngle}). Fig. \ref{fig:subfig:GEO-LEOlink_MTTE_TTP} only presents the transmission power of GEOS 1 allocated to LEOS $1$ for $t_{\mathrm{s}}=0$s and the transmission powers for other LEOSs are nearly the same as that of LEOS 1. It can be seen that although the transmission power obtained by the proposed allocation algorithm (denoted by ``Opt") can be higher at some time points than the constant power allocation (denoted by ``Sub-opt"), the nonzero power values are distributed along a shorter continuous time period and are all below the maximum beam power constraint $P_{\max}=40$W. The reason for some zero points at the beginning for the constant power allocation is that the starting link time of LEOS 1 $t_{\mathrm{s},1}$ is later than $t_{\mathrm{s}}=0$s and LEOS 5 establishes the link first to GEOS 1, i.e., $t_{\mathrm{s},5}=t_{\mathrm{s}}=0$s. Moreover, when we calculate the transmission energy of GEOS 1 on each LEOS and present them in Fig. \ref{fig:subfig:GEO-LEOlink_MTTE_TE}, it can be seen that the proposed transmission power allocation consumes lower transmission energy for each GEOS-LEOS downlink than that of the constant power allocation strategy. The main cause of the largest energy consumption for LEOS $5$ is the smallest channel link gain from GEOS 1 to LEOS 5.
\par
\begin{figure}[ht!]
   \subfigure[]
  {\begin{minipage}{1\textwidth}
 \centering
  \includegraphics[scale=0.8]{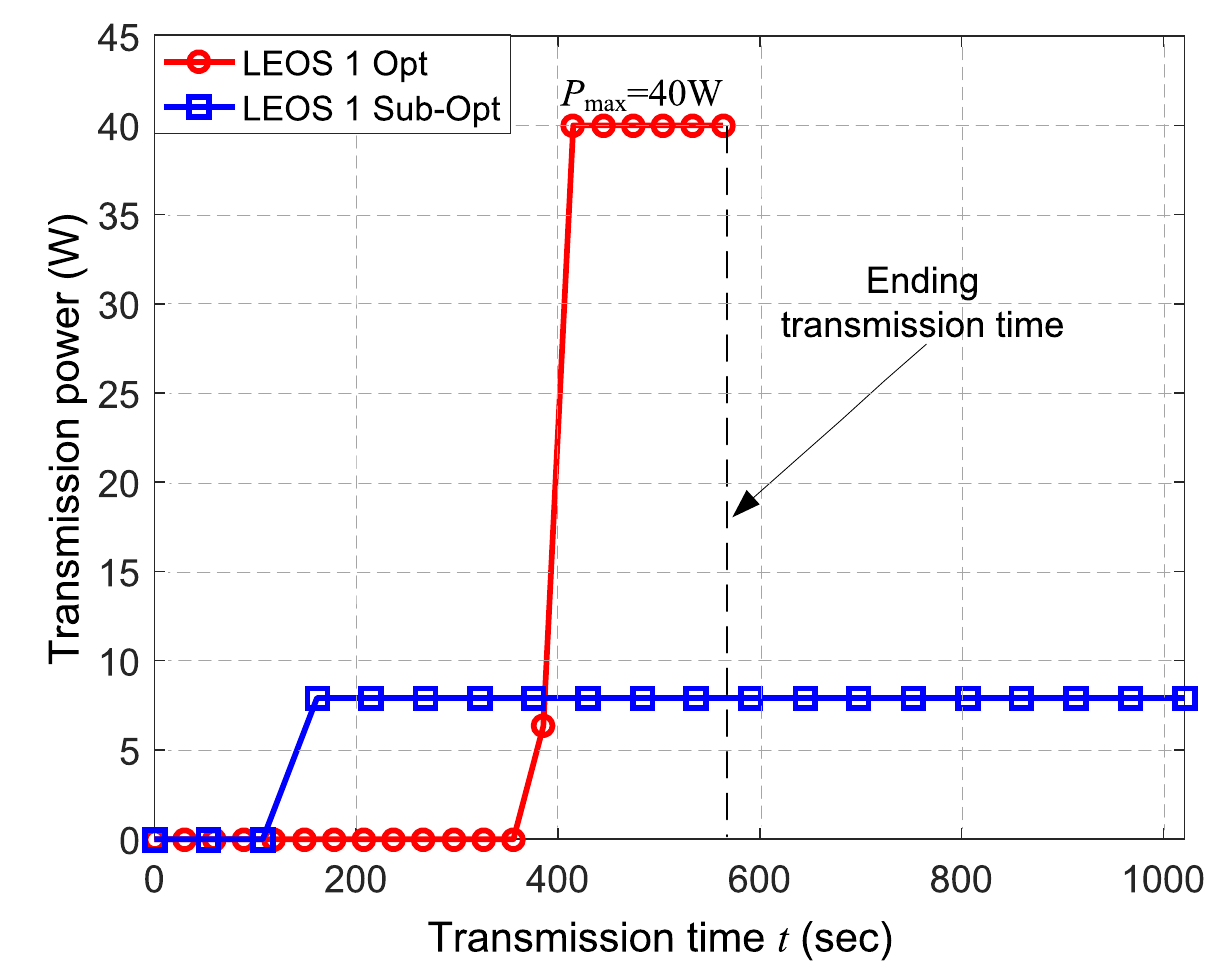}\label{fig:subfig:GEO-LEOlink_METT_ITP}
  \end{minipage}}\\
   \subfigure[]
  {\begin{minipage}{1\textwidth}
  \centering
  \includegraphics[scale=0.8]{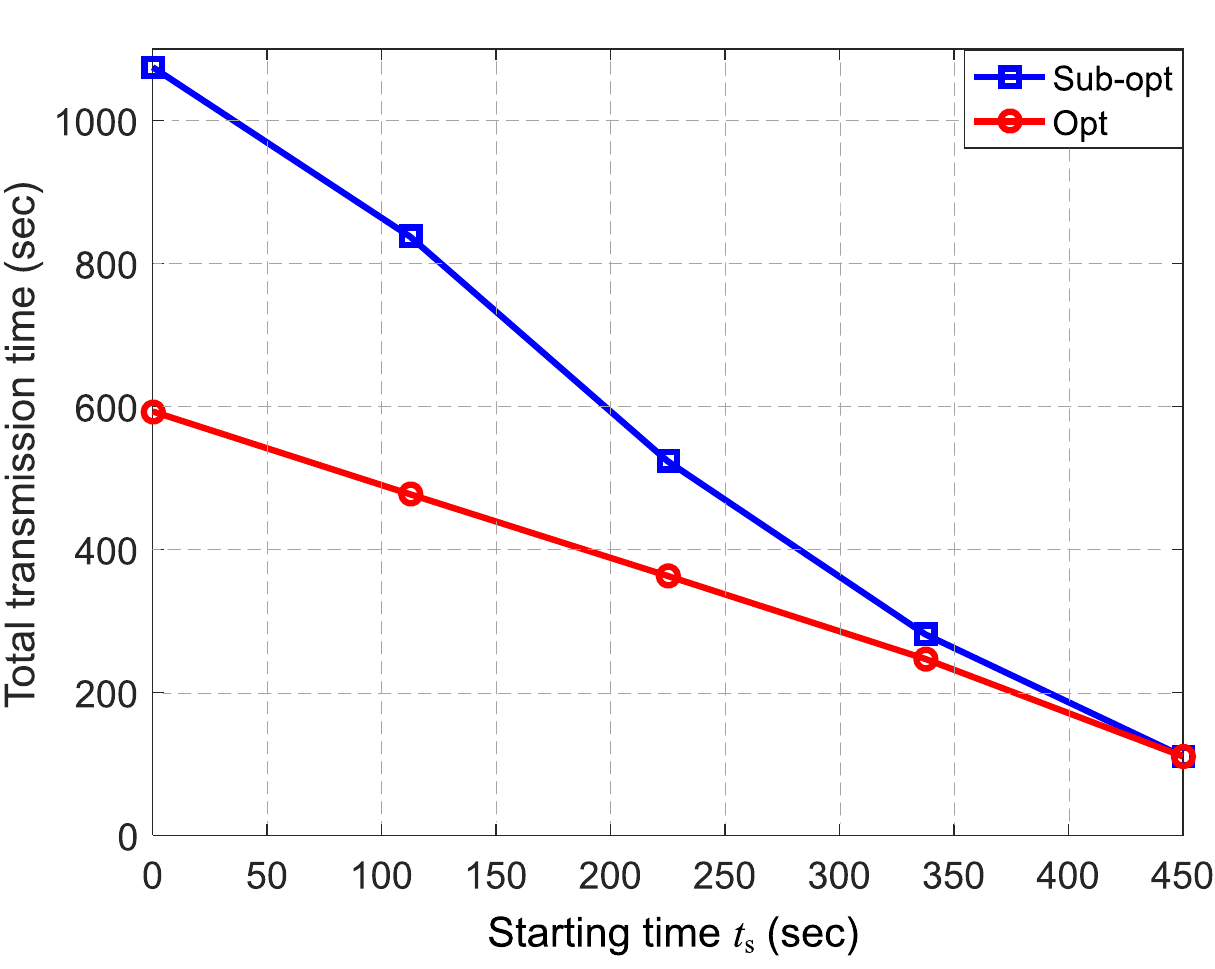}\label{fig:subfig:GEO-LEOlink_METT_TTE}
  \end{minipage}}
  \caption{Optimization results for minimizing total transmission time over the GEOS-LEOS downlink with $P_{\max}=40$W and $E_{\max}=3.7\times10^{4}$J. (a) Transmission power v.s. transmission time for starting time $t_{\mathrm{s}}=0$. (b) Total transmission time v.s. starting time $t_{\mathrm{s}}$ from $0$s to $450$s.}\label{fig:Fig_LEO-GEOlink_TimeMin_AlgorithmVerification}
  \end{figure}

  Next we compare the transmission power allocation (denoted by ``Opt") with the constant power allocation (denoted by ``Sub-opt") for minimizing the total transmission time. The maximum beam transmission power and total transmission energy budgets for GEOS 1 are set to be $P_{\max}=40$W and $E_{\max}=3.7\times10^{4}$J, respectively. The transmission time for the constant power allocation is obtained by solving problem (\ref{equ:Timeminimization_GEO-LEO}) under the assumption that the transmission power $P_{n}(t)$ for any $n\in\mathcal{N}$ is constant. Thus the corresponding problem can also be solved by using the same method as that for solving (\ref{equ:Timeminimization_GEO-LEO}). Fig. \ref{fig:subfig:GEO-LEOlink_METT_ITP} presents a comparison on the transmission power of LEOS 1 at the starting time $t_{\mathrm{s}}=0$s. Minimizing the total transmission time is actually equivalent to minimizing the ending transmission time of GEOS 1 since the starting time is assumed to be known. We can see that the proposed transmission power allocation shows earlier ending transmission time. The results of the total transmission time of the two algorithms for the starting time $t_{\mathrm{s}}$ ranging from $0$s to $600$s are presented in Fig.\ref{fig:subfig:GEO-LEOlink_METT_TTE}, from which we can still see that the proposed allocation shows shorter transmission time. And the main cause of the same transmission time at $t_{\mathrm{s}}=450$s for the two algorithms is the utilization of the same energy budget $E_{\max}$ for all the starting time points in the simulation, while $E_{\max}$ is superfluous for GEOS 1 at $t_{\mathrm{s}}=450$s to transmit $N\alpha$ files to the $N$ LEOSs, i.e., the actually consumed maximum transmission energy $E_{0}<E_{\max}$ and the common minimum transmission time is $T_{0}=\max\limits_{n\in \mathcal{N}} T_{n0}$ with $T_{n0}$ obtained by solving (\ref{equ:Timeminimization_GEO-LEO_critialpoint}). Note that the total transmission time decreases with the starting time since the channel gain of the whole GEOS-LEOS link increases with the starting time in our simulation scenario.
\subsection{Results for the LEOS-GEOS Uplink} \label{sec:SimResults:subsec:LEO-GEO}
\par
\begin{figure}[ht!]
   \subfigure[]
  {\begin{minipage}{1\textwidth}
  \centering
  \includegraphics[scale=0.8]{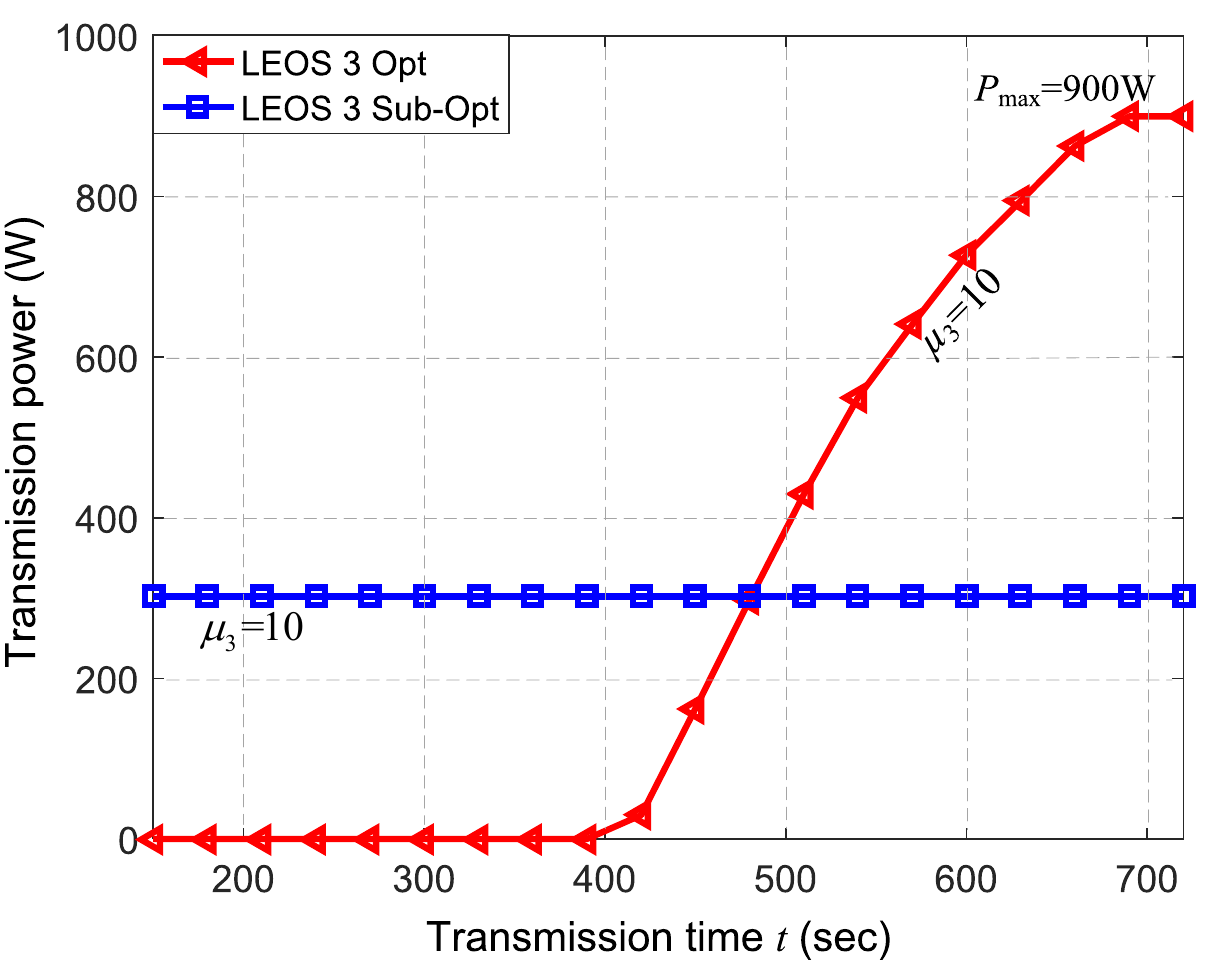}\label{fig:subfig:LEO-GEOlink_MTTE_TTP}
  \end{minipage}}\\
   \subfigure[]
  {\begin{minipage}{1\textwidth}
  \centering
  \includegraphics[scale=0.8]{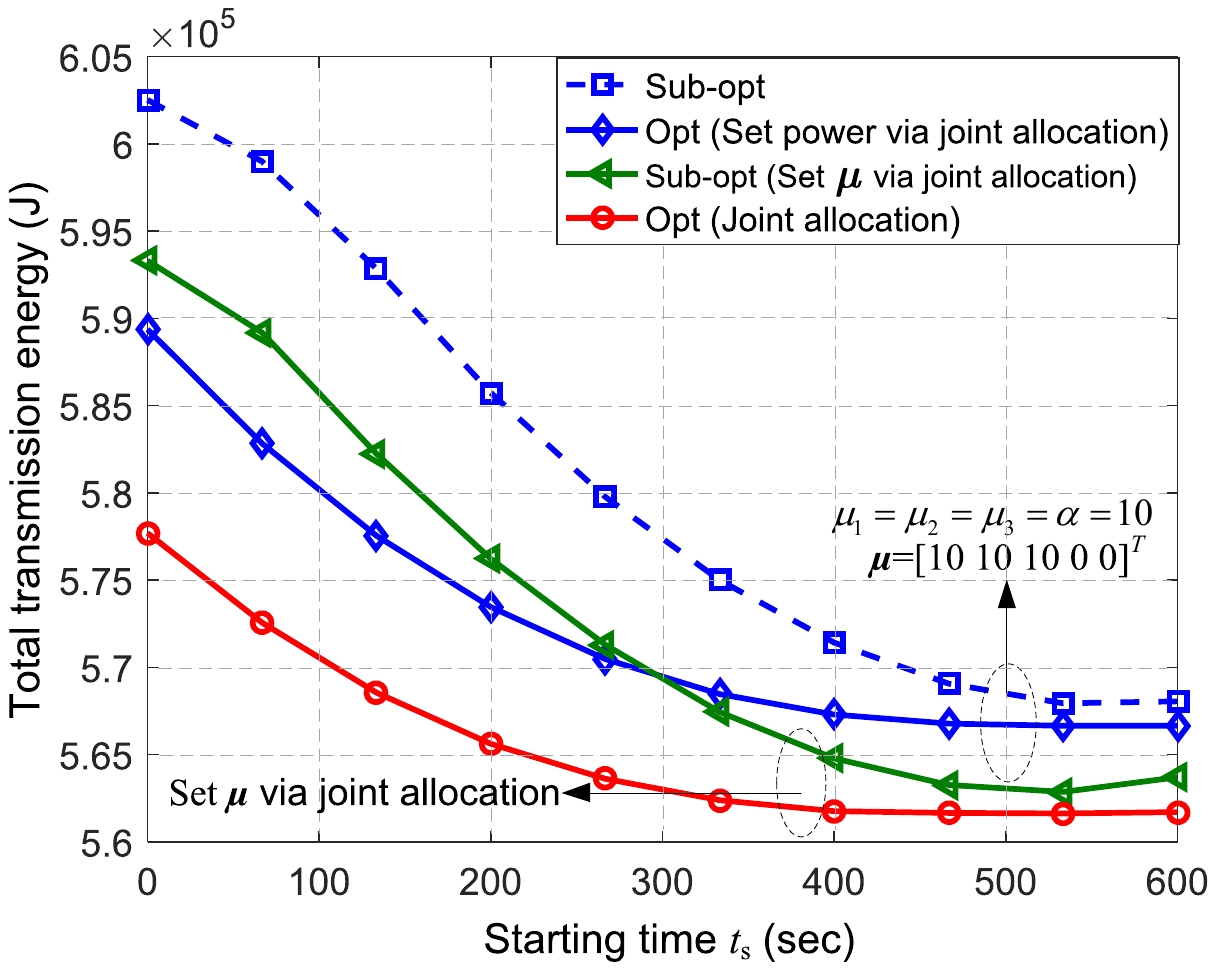}\label{fig:subfig:LEO-GEOlink_MTTE_TTE}
  \end{minipage}}
  \caption{Optimization results for minimizing total transmission energy over the LEOS-GEOS uplink with $T=450$s and $P_{\max}=900$W. (a) Transmission power v.s. transmission time for $t_{\mathrm{s}}=133$s. (b) Total transmission energy v.s. starting time $t_{\mathrm{s}}$ from $0$s to $600$s.}\label{fig:Fig_LEO-GEOlink_MTTE}
  \end{figure}

 We compare the joint uploaded-data size and power allocation to minimize the total transmission energy with the constant power allocation. In the following simulation results, we use ``Opt" to represent the allocations containing transmission power allocation while use ``Sub-opt" to represent the allocations using constant power. Similarly to the simulations performed in Section \ref{sec:SimResults:subsec:GEO-LEO}, we set $t_{\mathrm{s},n}=\max\{t_{\mathrm{s}},t_{0n}\}$ and $t_{\mathrm{e},n}=t_{\mathrm{s}}+T$ for the $N=5$ LEOSs and let $T=600$s. We assume that the LEOSs will transmit totally 30 files directly under the $(\alpha,K)$-reconstructability by $\mu_{1}=\mu_{2}=\mu_{3}=\alpha=10$; while for the joint uploaded-data size and power allocation, we can obtain the optimal file numbers transmitted by the five LEOSs via optimization for any given $t_{\mathrm{s}}$. For example, when $t_{\mathrm{s}}=133$s, the obtained $\bm\mu=[0~5~10~10~5]^{T}$, which implies that LEOSs 2-5 need to transmit, and for simplicity, Fig. \ref{fig:subfig:LEO-GEOlink_MTTE_TTP} only presents the transmission power of LEOS $3$ over the time interval $T$ and the transmission powers of other LEOSs that transmit data behave similarly. We can see from the figure that the joint uploaded-data size and power allocation needs a shorter continuous transmission time period. Although it may need higher instantaneous transmission power at certain time, all the higher values are below the maximum power constraint $P_{\max}=900$W. To analyze the total transmission energy, four transmission strategies are considered: constant transmission power without resource allocation by letting $\bm\mu=[10~10~10~0~0]^{T}$, transmission power allocation based on $\bm\mu=[10~10~10~0~0]^{T}$, constant transmission power based on $\bm\mu$ obtained by the joint uploaded-data size and power allocation, and directly using the joint uploaded-data size and power allocation. The results with respect to $t_{\mathrm{s}}$ from $0$s to $600$s are presented in Fig. \ref{fig:subfig:LEO-GEOlink_MTTE_TTE}, where the constant transmission power without resource allocation, which can be viewed as $(\alpha,K)$-reconstructability, achieves the highest total transmission energy since it does not include any optimization while the joint uploaded-data size and power allocation achieves the lowest total transmission energy since it includes downloaded-file size and transmission power allocations.
  \par
\begin{figure}[ht!]
  {\begin{minipage}{1\textwidth}
 \centering
  \includegraphics[scale=0.8]{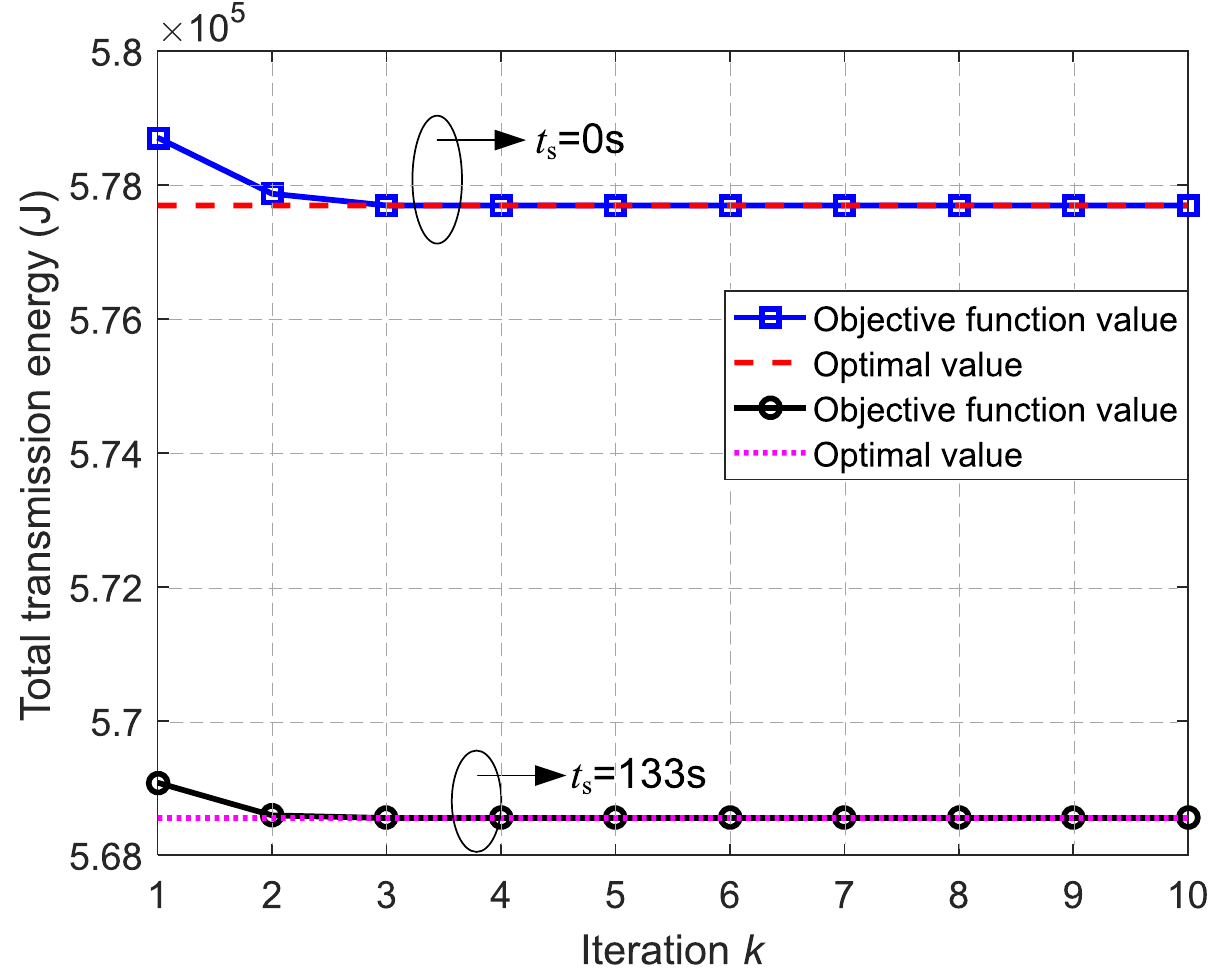}
  \end{minipage}}\\
  \caption{Convergence of the OA method used in \emph{Algorithm} \ref{alg:OA-DSA}.}\label{fig:subfig:GEO-LEOlink_ConvergenceAnalysis}
  \end{figure}

 As both \emph{Algorithms} \ref{alg:OA-DSA} and \ref{alg:LEO-GEO_MiniTime} employ the OA method which is an iterative procedure, we show its convergence behavior in Fig. \ref{fig:subfig:GEO-LEOlink_ConvergenceAnalysis}, where the values of the objective function in (\ref{equ:Powerminimization_Reconst_Eq1}) is plotted against the number of iterations when $t_{\mathrm{s}}=0$s and $t_{\mathrm{s}}=133$s, respectively. The optimal values are obtained by using the constrained waterfilling algorithm according to the obtained optimal $\bm\mu$. From the figure we can see that for both cases 2 or 3 iterations suffice to obtain the optimal solutions.
\par
\begin{figure}[ht!]
   \subfigure[]
  {\begin{minipage}{1\textwidth}
  \centering
  \includegraphics[scale=0.8]{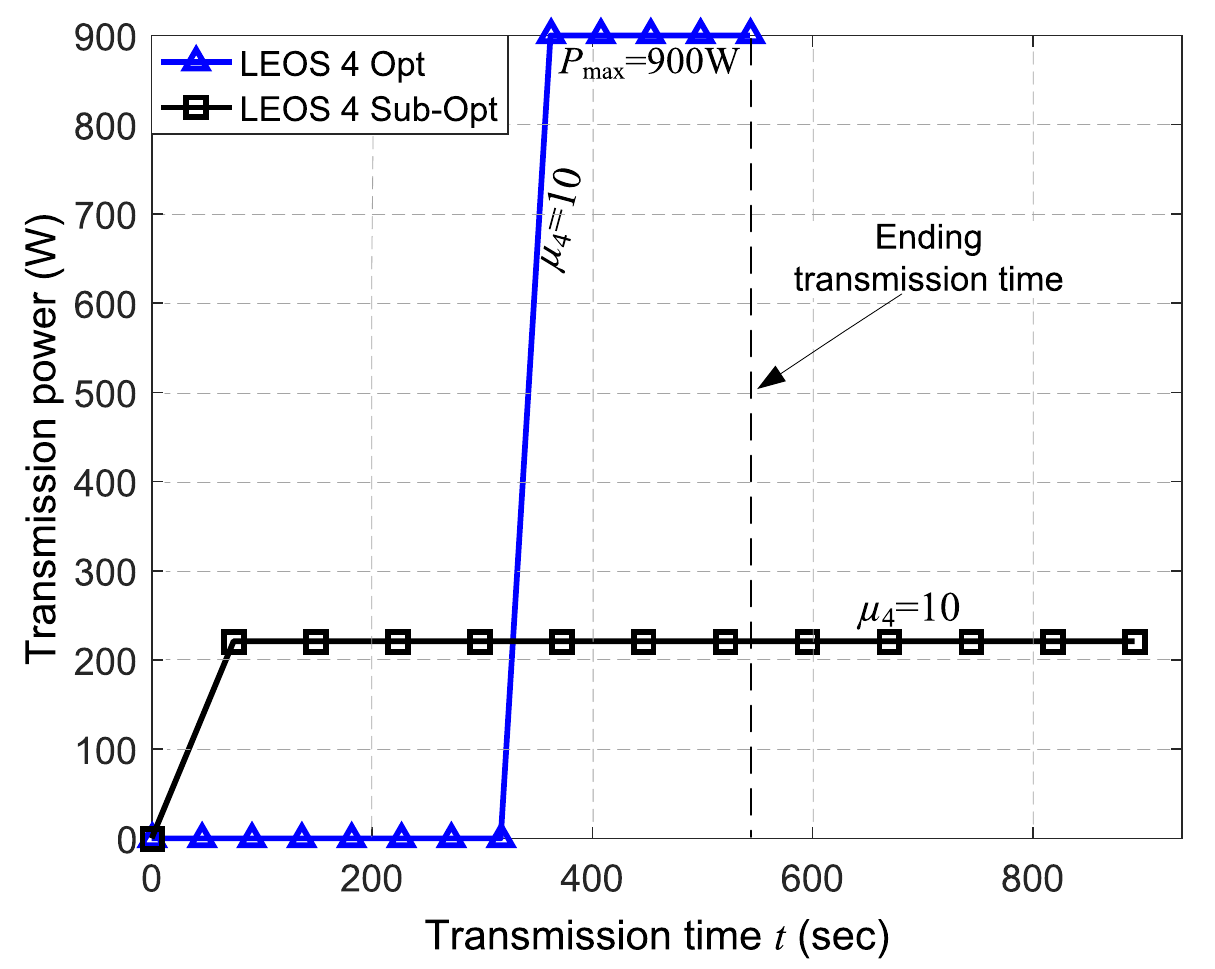}\label{fig:subfig:LEO-GEOlink_MWRT_a}
  \end{minipage}}\\
    \subfigure[]
   {\begin{minipage}{1\textwidth}
   \centering
   \includegraphics[scale=0.8]{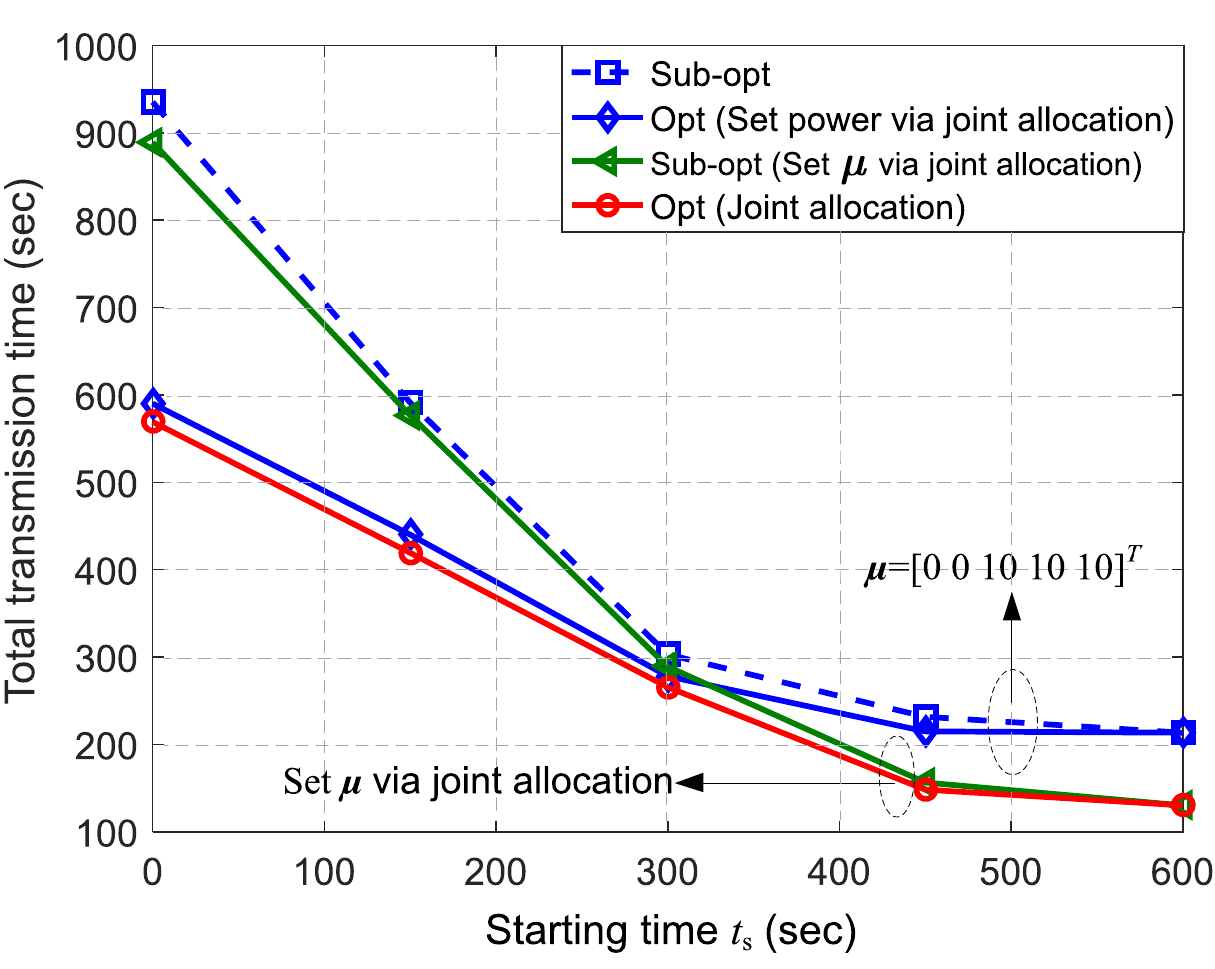}\label{fig:subfig:LEO-GEOlink_MWRT_b}
   \end{minipage}}
  \caption{Optimization results for minimizing total transmission time over the LEOS-GEOS uplink with $P_{\max}=900$W and $E_{\max}=5.8\times10^{5}$J. (a) Transmission power v.s. transmission time for starting time $t_{\mathrm{s}}=0$. (b) Total transmission time v.s. starting time $t_{\mathrm{s}}$ from $0$s to $600$s.}\label{fig:Fig_LEO-GEOlink_MWRT}
  \end{figure}

 Next we compare the joint uploaded-data size and power allocation to minimize the total transmission time with the constant power allocation. Set the total transmission energy budget $E_{\max}=5.8\times10^{5}$J and let the other parameters be the same with the forgoing analyses. We assume that the LEOSs will transmit totally 30 files directly by $\mu_{3}=\mu_{4}=\mu_{5}=\alpha=10$ for the constant power allocation; while for the joint allocation, the obtained optimal file numbers transmitted by the five LEOSs for $t_{\mathrm{s}}=0$s are $\bm\mu=[0~5~9~10~6]^{T}$. The results of the transmission power of LEOS 4 for the two allocations are presented in Fig. \ref{fig:subfig:LEO-GEOlink_MWRT_a}. As minimizing the total transmission time is actually equivalent to minimizing the ending transmission time of the LEOS network since the starting time is assumed to be known, we can see from the figure that the joint allocation just needs a shorter transmission time and all the values of the transmission power are below $P_{\max}=900$W. To further analyze the total transmission time, two additional allocations are also considered: minimizing time via solving problem (\ref{equ:Timeminimization_LEO-GEO}) given that $\bm\mu=[0~0~10~10~10]^{T}$ and minimizing time via solving problem (\ref{equ:Timeminimization_LEO-GEO}) given that $P_{n}(t)$ for all $n\in\mathcal{N}$ are constant. Since the subsequent two allocations are the special cases of the joint allocation given by (\ref{equ:Timeminimization_LEO-GEO}), both of them can be easily solved and we omit the details of their solving procedures. The minimum total transmission times of the four allocations with respect to $t_{\mathrm{s}}$ from $0$s to $600$s are presented in Fig.\ref{fig:subfig:LEO-GEOlink_MWRT_b}, from which we can see that the joint uploaded-data size and power allocation shows the shortest total transmission time while the constant power allocation under $\bm\mu=[0~0~10~10~10]^{T}$ shows the longest transmission time and the results of the other two allocations line between of the two. This is because the constant power allocation under $\bm\mu=[0~0~10~10~10]^{T}$ (denoted by ``Sub-opt" in Fig. \ref{fig:subfig:LEO-GEOlink_MWRT_b}) just takes the maximum individual transmission time of the LEOSs as the common minimum transmission time under the energy budget constraint and does not optimize the power allocation over time.
  \subsection{Results for Failed LEOS Repair}
\begin{figure}[ht!]
   \subfigure[]
  {\begin{minipage}{1\textwidth}
  \centering
  \includegraphics[scale=0.8]{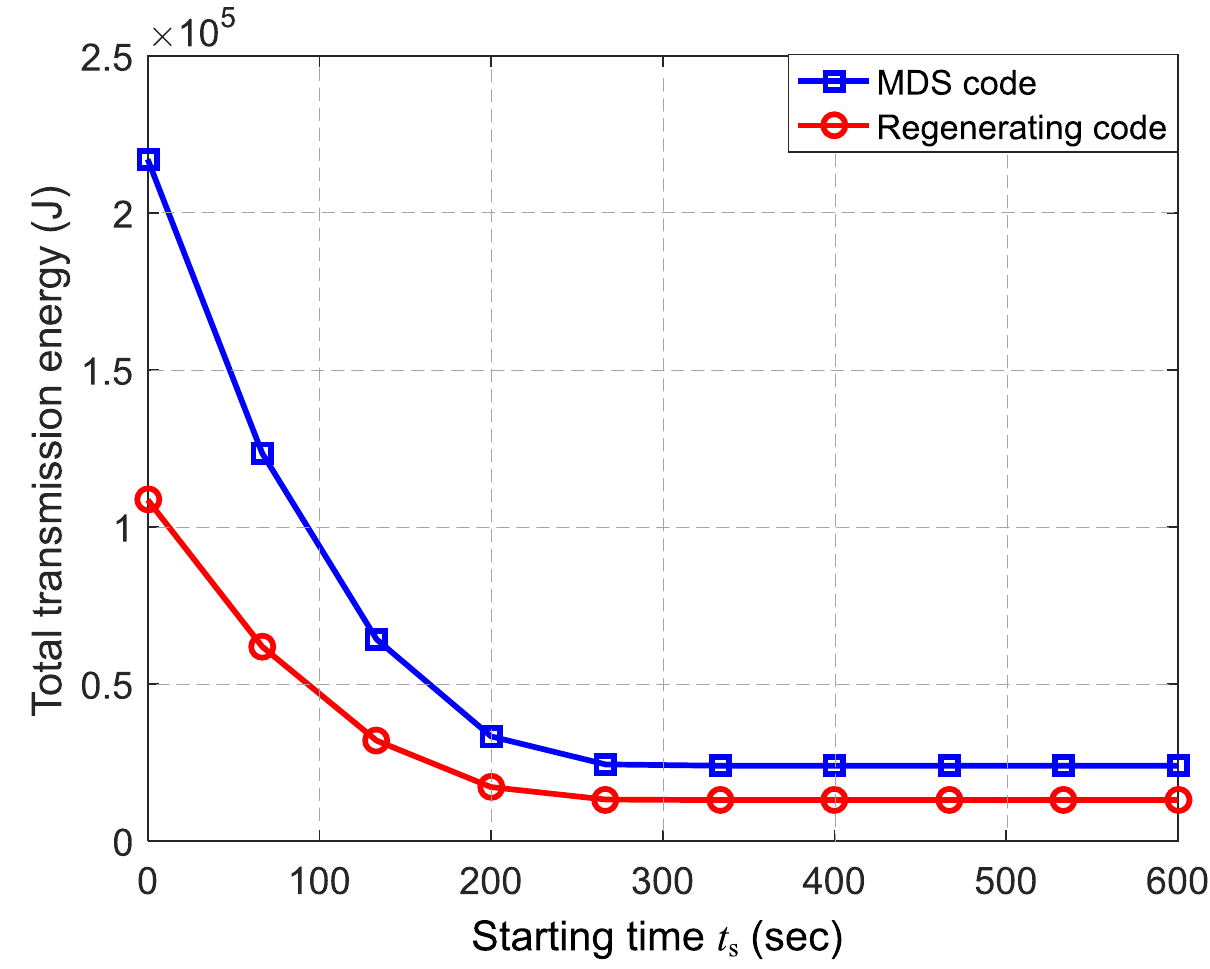}\label{fig:subfig:LEOSRepair_MTTE}
  \end{minipage}}\\
    \subfigure[]
   {\begin{minipage}{1\textwidth}
   \centering
   \includegraphics[scale=0.8]{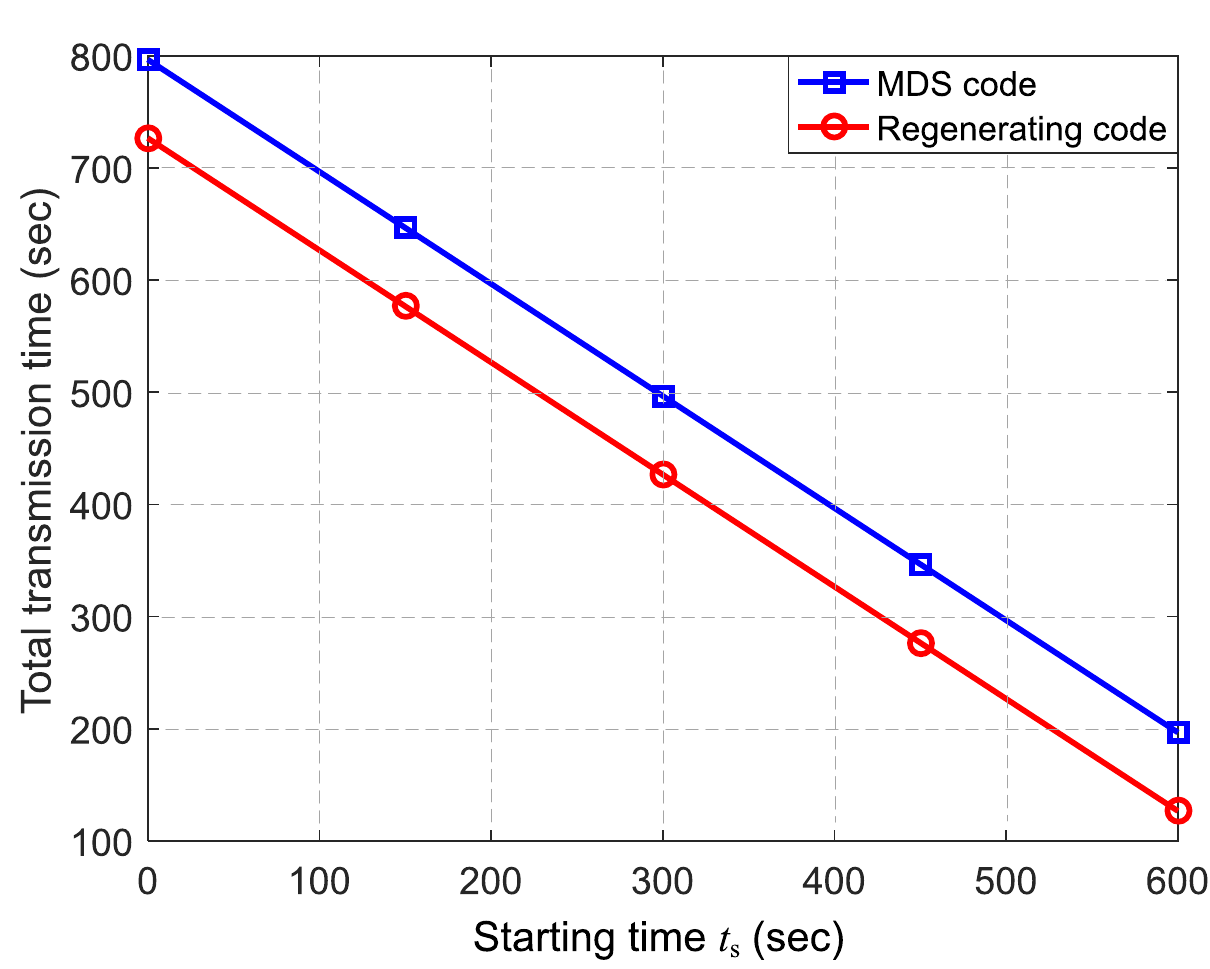}\label{fig:subfig:LEOSRepair_MTTT}
   \end{minipage}}
  \caption{Optimization results for failed LEOS repair with $P_{\max}=900$W. (a) Optimal total transmission energy v.s. starting time $t_{\mathrm{s}}$ from $0$s to $600$s. (b) Optimal total transmission time v.s. starting time $t_{\mathrm{s}}$ from $0$s to $600$s, where $E_{\max}=3.1\times10^{4}$J.}\label{fig:Fig3_LEOSRepair}
  \end{figure}

We present some results for the resource allocation to minimize total transmission energy and time between the regenerating and MDS codes. Assume that the data in LEOS 5 are failed. Then LEOS $5$ needs to download some data from other LEOSs to regenerate its $\alpha$ files. According to the simulation parameters given in Table \ref{tab:Params_SimRes}, we have $D=4$, $\beta=5$, which means LEOS $5$ needs to choose other $D=4$ LEOSs and download $\beta = 5$ files from each of them under the regenerating code. Note that under the MDS code, the amount of downloaded data to repair a faild node should be no less than the amount of the original data. Thus LEOS $5$ should download at least 30 files when using the MDS code. Obviously we can see that the regenerating code can have higher efficiency than the MDS code in failed node repair since the former only needs to download $\gamma=D\beta=20$ files. For simplicity, assume all the LEOSs move in an orbital-plane. Then the instantaneous distance between any two LEOSs can be easily calculated based on the other orbital parameters given in Table \ref{tab:Params_SimRes}. Thus we can formulate some similar optimization problems to obtain the minimum total transmission energy and time, respectively. The corresponding results are presented in Fig. \ref{fig:Fig3_LEOSRepair}, which shows the higher efficiency of the regenerating code in repairing LEOS $5$.
\section{Conclusions}\label{sec:Conclusion}
In this paper, we have considered relay-assisted inter-GEOSs communications through LEOSs using distributed-storage coding. We have proposed and optimized the transmission schemes for both the GEOS-LEOS downlink and the LEOS-GEOS uplink. For the GEOS-LEOS downlink, a regenerating-code-based transmission scheme has been optimized to guarantee data reconstructability, and to minimize the total transmission energy or time. For the LEOS-GEOS uplink, we have proposed a flexible partial-downloading coding transmission scheme to guarantee data reconstructability, and have considered the problem of joint uploaded-data size and power allocation to minimize the total transmission energy or time. Extensive simulation results have been presented to show the advantages of the proposed algorithms.

\appendices
\section{Derivations of the OA Algorithm for Problem (\ref{equ:Powerminimization_Reconst_Eq1}) }\label{app:OAmethod_LEC-INLP}
\par The iterative computation in the OA method is to initialize and improve the bounds of the obtained solutions over feasible region. The lower bounds are obtained by relaxing the original problems while the upper bounds are obtained by finding feasible points. To solve the MINLP formulated in (\ref{equ:Powerminimization_Reconst_Eq1}), three problems will be considered. The first problem is the continuous relaxation of the MINLP given by
\begin{equation}\label{equ:PowerminimizationLEO-GEO_Relaxation_NLPR}
\mathrm{NLPR}\left\{\begin{array}{cl}
    \min\limits_{\mathcal{P},\bm\mu} &f_{\mathrm{E}}(\mathcal{P})\\
      \mathrm{s.t.}&\displaystyle\sum\limits_{n\in \mathcal{N}} \mu_{n}=M,~0\leq\mu_{n}\leq\alpha,~\forall n\in\mathcal{N},\\
      &\displaystyle\begin{aligned}
      & f_{n}(\mathfrak{p}_{n},\mu_{n})\leq 0,~\forall n\in\mathcal{N},\\
      & 0\leq \mathcal{P}\leq P_{\max},
      \end{aligned}
  \end{array}\right.
\end{equation}
which can be solved by a convex solver and lead to a lower bound of problem (\ref{equ:Powerminimization_Reconst_Eq1}).
\par  Letting $\widehat{\bm\mu}$ be a feasible point of problem (\ref{equ:Powerminimization_Reconst_Eq1}), we can obtain the second problem formulated as
\begin{equation}\label{equ:PowerminimizationLEO-GEO_NLP}
{\mathrm{NLP}}\left\{ \begin{array}{cl}
    \displaystyle\min\limits_{\mathcal{P}} &\displaystyle f_{\mathrm{E}}(\mathcal{P})\\
     \displaystyle\mathrm{s.t.}  & \displaystyle f_{n}(\mathfrak{p}_{n},\widehat{\mu}_{n})\leq 0,~\forall n\in\mathcal{N},\\
      &\displaystyle 0\leq \mathcal{P}\leq P_{\max},
\end{array} \right.
\end{equation}
which can be directly solved by a convex solver and is used to determine the upper bound on $f_{\mathrm{E}}(\mathcal{P})$ in each iteration.
\par Initialize the OA-point set $\mathcal{S}_{\mathcal{P},\bm\mu}$ by solving (\ref{equ:PowerminimizationLEO-GEO_Relaxation_NLPR}) and repeatedly solve the relaxed master problem of (\ref{equ:Powerminimization_Reconst_Eq1}), which is the third problem obtained by replacing the nonlinear constraints by their linear outer approximations at the points of set $\mathcal{S}_{\mathcal{P},\bm\mu}$. The corresponding master problem is given by
\begin{equation}\label{equ:PowerminimizationLEO-GEO_OA-ILP}
   \mathrm{OA}\mathrm{-}\mathrm{ILP}\left\{
   \begin{array}{cl}
    \displaystyle\min\limits_{\mathcal{P},\bm\mu} &\displaystyle f_{\mathrm{E}}(\mathcal{P})\\
      \displaystyle\mathrm{s.t.}&\displaystyle f_{n}(\overline{\mathfrak{p}}_{n},\overline{\mu}_{n})+\bm\nabla^{T} f_{n}(\overline{\mathfrak{p}}_{n},\overline{\mu}_{n})\left(
\begin{array}{c}
  \mathfrak{p}_{n} - \overline{\mathfrak{p}}_{n}\\
\mu_{n} - \overline{\mu}_{n}
\end{array}
 \right)\leq 0,~\forall(\overline{\mathfrak{p}}_{n},\overline{\mu}_{n})\in \mathcal{S}_{\mathcal{P},\bm\mu},\\
      &\displaystyle\sum\limits_{n\in \mathcal{N}} \mu_{n}=M,~\mu_{n}\in\{0,1,\cdots,\alpha\},~\forall n\in\mathcal{N},
  \end{array}\right.
\end{equation}
which is a mixed integer linear program (MILP) and can be readily solved by many efficient branch-and-cut-based linear programming solvers such as CPLEX, LINDO, INTLINPROG, etc.\cite{Mitchell2002,Linderoth2010}. In each iteration, $\mathcal{S}_{\mathcal{P},\bm\mu}$ is updated by collecting $(\mathcal{P}_{\mathrm{NLP}},\bm\mu_{\mathrm{OA}})$, where $\mathcal{P}_{\mathrm{NLP}}$ is the optimal solution to (\ref{equ:PowerminimizationLEO-GEO_NLP}) under $\widehat{\bm\mu}=\bm\mu_{\mathrm{OA}}$ while $\bm\mu_{\mathrm{OA}}$ is the optimal solution to (\ref{equ:PowerminimizationLEO-GEO_OA-ILP}), and the lower bound on $f_{\mathrm{E}}(\mathcal{P})$ is updated by the optimal value of (\ref{equ:PowerminimizationLEO-GEO_OA-ILP}).

\ifCLASSOPTIONcaptionsoff
  \newpage
\fi

\bibliographystyle{IEEEtran}
\end{document}